\documentclass[onecolumn,11pt,showpacs,notitlepage,nofootinbib,floatfix,superscriptaddress]{revtex4-1} 
\usepackage{graphicx}
\usepackage{amssymb}
\usepackage{amsmath}
\usepackage{amsthm}
\usepackage{amsfonts}
\usepackage{mathtools}
\usepackage{xcolor}
\usepackage{hyperref}
\usepackage{soul}
\usepackage{csquotes}
\usepackage{multirow}
\usepackage{hhline}
\usepackage{siunitx}
\usepackage{verbatim}

\newcommand{\eq}[1]{\hyperref[eq:#1]{Eq.~(\ref*{eq:#1})}}
\renewcommand{\sec}[1]{\hyperref[sec:#1]{Section~\ref*{sec:#1}}}
\newcommand{\secsm}[1]{\hyperref[sec:#1]{Sec.~\ref*{sec:#1}}}
\newcommand{\app}[1]{\hyperref[app:#1]{Appendix~\ref*{app:#1}}}
\newcommand{\theo}[1]{\hyperref[thm:#1]{Theorem~\ref*{thm:#1}}}
\newcommand{\algo}[1]{\hyperref[alg:#1]{Algorithm~\ref*{alg:#1}}}
\newcommand{\lemm}[1]{\hyperref[lem:#1]{Lemma~\ref*{lem:#1}}}
\newcommand{\defn}[1]{\hyperref[defn:#1]{Definition~\ref*{defn:#1}}}
\newcommand{\corr}[1]{\hyperref[cor:#1]{Corollary~\ref*{cor:#1}}}
\newcommand{\fig}[1]{\hyperref[fig:#1]{Fig.~\ref*{fig:#1}}}
\newcommand{\tab}[1]{\hyperref[tab:#1]{Table~\ref*{tab:#1}}}
\newcommand{\tabsm}[1]{\hyperref[tab:#1]{Tab.~\ref*{tab:#1}}}
\newcommand{\propos}[1]{\hyperref[prop:#1]{Proposition~\ref*{prop:#1}}}
\newcommand{\propsm}[1]{\hyperref[prop:#1]{Prop.~\ref*{prop:#1}}}
\newcommand{\rema}[1]{\hyperref[rem:#1]{Remark~\ref*{rem:#1}}}

\DeclareMathOperator*{\argmin}{arg\,min}

\newcommand{\ket}[1]{|#1\rangle}
\newcommand{\bra}[1]{\langle #1|}

\newcommand{\Q}{Q}
\newcommand{\Qalg}{Q_{\text{alg}}}
\newcommand{\TimeLogical}{C}
\newcommand{\TimeLogicalFast}{\TimeLogical_{\text{min}}}
\newcommand{\R}{M}
\newcommand{\RTof}{M_{\text{Tof}}}
\newcommand{\RT}{M_{T}}
\newcommand{\RMeas}{M_\text{meas}}
\newcommand{\RU}{M_R}
\newcommand{\depthU}{D_R}

\newcommand{\Plog}{P}
\newcommand{\PlogDis}{P_T}

\newcommand{\tphy}{t}
\newcommand{\qphy}{q}
\newcommand{\TperU}{R_T}
\newcommand{\tGate}{t_\text{gate}}
\newcommand{\tMeas}{t_\text{meas}}

\newcommand{\nFactories}{F}
\newcommand{\epsSynth}{\epsilon_{\text{syn}}}
\newcommand{\epsLog}{\epsilon_{\text{log}}}
\newcommand{\epsDis}{\epsilon_{\text{dis}}}
\newcommand{\epsTot}{\epsilon}

\begin{document}

\title{
Assessing requirements to scale to practical quantum advantage
}

\newcommand{\ethz}		{ETH Zurich, Department of Computer Science, Zürich, 8006, Switzerland}
\newcommand{\redmond}		{Microsoft Quantum, Redmond, WA 98052, USA}
\newcommand{\santabarbara}	{Station Q, Microsoft Quantum, Santa Barbara, California 93106, USA}
\newcommand{\zurich}			{Microsoft Quantum, Zurich, Switzerland}
\newcommand{\copenhagen}		{Microsoft Quantum Lab Copenhagen, 2100 Copenhagen, Denmark}
\newcommand{\delft}			{Microsoft Quantum Lab Delft, 2600 GA Delft, The Netherlands}
\newcommand{\sherbrooke}		{Microsoft Quantum, Sherbrooke, QC, Canada}
\newcommand{\sydney}			{Microsoft Quantum, The University of Sydney, NSW, 2006, Australia}

\author{M.~E.~Beverland}
\affiliation{\redmond}
\author{P.~Murali}
\affiliation{\redmond}
\author{M.~Troyer}
\affiliation{\redmond}
\author{K.~M.~Svore}
\affiliation{\redmond}
\author{T.~Hoefler}
\affiliation{\ethz}
\author{V.~Kliuchnikov}
\affiliation{\redmond}
\author{G.~H.~Low}
\affiliation{\redmond}
\author{M.~Soeken}
\affiliation{\zurich}
\author{A.~Sundaram}
\affiliation{\redmond}
\author{A.~Vaschillo}
\affiliation{\redmond}
\date{\today}

\begin{abstract}
While quantum computers promise to solve some scientifically and commercially valuable problems thought intractable for classical machines, delivering on this promise will require a \textit{large-scale} quantum machine. 
Understanding the impact of architecture design choices for a scaled quantum stack for specific applications, prior to full realization of the quantum system, is an important open challenge.  
To this end, we develop a framework for quantum resource estimation, abstracting the layers of the stack, to estimate resources required across these layers for large-scale quantum applications. 
Using a tool that implements this framework, we assess three scaled quantum applications and find that hundreds of thousands to millions of physical qubits are needed to achieve practical quantum advantage. 
We identify three qubit parameters, namely size, speed, and controllability, that are critical at scale to rendering these applications practical. 
A goal of our work is to accelerate progress towards practical quantum advantage by enabling the broader community to explore design choices across the stack, from algorithms to qubits.
\end{abstract}
\maketitle

\section{Toward quantum applications with practical impact}
With rapid progress in quantum computing, there is a shift in focus from scientific demonstrations on a small number of noisy qubits \cite{arute2019,Preskill2018NISQ} towards large quantum systems solving valuable business problems and resolving long-standing scientific questions that are classically intractable~\cite{Schoelkopf21}.
But just how large does a quantum computer need to be to achieve these forms of \textit{practical quantum advantage},
and how long will such a computation take?
Are some qubit technologies better suited than others for solving such problems?
What are the best architecture choices across the hardware and software stacks to enable scaled quantum computation?

While many quantum algorithms have asymptotic speedups over their non-quantum counterparts~\cite{jordan2011}, scaling up to problem sizes with a practical quantum advantage \cite{Babbush2020} will demand high-precision quantum computers.
Quantum operations at the physical level are noisy, and so the long computations required for practical quantum advantage necessarily require error correction to achieve fault tolerance.
Quantum computers are complex, and the overheads required to ensure fault-tolerant quantum computing will significantly outpace the resources required for fault-tolerant classical computing which is based on small, cheap, and reliable transistors. 
A fault-tolerant operation on a quantum computer requires orders of magnitude more space -- including many transistors for qubit control and readout -- and runs with much slower clock speeds than a classical computer.
With these overheads, practical quantum advantage will be achieved, albeit only for algorithms with small I/O requirements and superquadratic (ideally exponential) speedups over their classical counterparts \cite{Love2020,Babbush2020,TroyerQ2B2020}.


One algorithm with superquadratic speedup for which the cost of error correction is well studied is Shor's period finding algorithm \cite{Shor1994}. 
Estimating the resources required for Shor's algorithm is important for assessing the vulnerability of some of today's public key cryptosystems to future quantum threats.
With the fastest quantum hardware operations proposed to date, factoring a 2048-bit integer using Shor's algorithm could be done in minutes with an array of twenty five thousand \textit{perfect, noiseless} qubits. 
Yet in reality, qubits are noisy and must have error correction to enable long computation, and so as we will reconfirm later, the implementation cost increases to about a day with tens of millions of qubits~\cite{Gidney2021}.

Some of the most compelling quantum algorithms with scientific and commercial interest, however, are those which leverage the ability of a quantum computer to efficiently simulate quantum systems, with applications across chemistry, materials science, condensed matter, and nuclear physics. 
The exact simulation time of the dynamics of such quantum systems scales exponentially with classical algorithms, but has a favorable polynomial scaling for quantum algorithms~\cite{lloyd1996}. 
The earliest application of scientific interest may be simulating the dynamics of around one hundred quantum spins in a quantum magnet \cite{daley2022}. 
The earliest commercially relevant applications will likely be quantum simulations of chemistry and materials science problems, where a quantum-accelerated elucidation of catalytic reaction mechanisms has applications to fertilizer production \cite{Reiher2017}, carbon fixation \cite{vonBurg2021}, among many other problems~\cite{bauer2020quantum,vonBurg2021,delgado2022,kim2022}. 

We propose a framework to understand the requirements of applications promising quantum advantage, determine their practicality, and assess changes in the underlying architecture to accelerate their implementation.
The framework is composed of layers of abstractions capable of estimating the performance and required resources of quantum applications and algorithms on current and future quantum architecture designs. 
Starting with an algorithm implemented in a high-level programming language, the framework compiles the algorithm onto the architecture of a specified quantum device and models the resource requirements at the architecture level, considering qubit layout, allowed operations, and other design parameters. 
We specify models for each layer and implement a tool, the Azure Quantum Resource Estimator, that calculates based on those models an estimate of the resources required to implement a given quantum algorithm on a given architecture with an underlying qubit technology. 
We focus on \emph{digital} quantum computers using gates and measurements for computation,
rather than analog quantum simulators and annealers \cite{daley2022};  however, similar layers of abstraction may also be valuable to analyze these other systems.  

We analyze the resources that would be required to implement three specific high-impact applications on various quantum architectures.
The first considers the \emph{quantum dynamics} of a simple quantum magnet, the so-called two-dimensional (2D) transverse field Ising model, where we consider a parameter regime that is on the boundary of what can be done by classical computation~\cite{Pearson2020}. 
The second is for \emph{quantum chemistry} where we analyze the activation energy of a catalyst for carbon fixation~\cite{vonBurg2021}.
Lastly, we analyze \emph{factoring} a large integer with Shor's algorithm.  
While we do not foresee running this application in practice, it is arguably the best-studied quantum algorithm and can help provide security requirements of our classical cryptosystems.
Our analysis reveals that practical quantum advantage for scientific and commercial problems requires hundreds of thousands to millions of qubits which satisfy the following requirements.

\vspace{0.5cm}
{\bf Requirements for scale.---}
To achieve practical quantum advantage, quantum computers will require an underlying qubit technology that at scale is:
\begin{itemize}
	\item {\em Controllable:} Practical quantum error correction requires reliable control of more than a million well-connected qubits, with parallel operations that fail in under one part in a thousand.
	
	\item {\em Fast:} To achieve a practical runtime of one month or less, while targeting a physical qubit count of around one million, operations will need to be performed in under a microsecond. 
	
	\item {\em Small:} Scaling to a million and more qubits constrains the size of the qubit to tens of microns in diameter; this size is determined to avoid the complexity of coherent high-bandwidth quantum interconnects between qubits on different modules.

\end{itemize}

We see these requirements as necessary to scale to practical quantum advantage, and as valuable additions, for scale, to the criteria for building a quantum computer proposed by DiVincenzo in the year 2000~\cite{Divincenzo2000}.
They stem from incorporating research developments spanning the last two decades, insights across algorithms to qubits, and an empirical end-to-end resource analysis of several applications and qubit parameter settings. 
We hope they reveal new considerations and valuable insights in considering how to engineer and architect for scale.

\section{Resource estimation for classical and quantum computing}

Estimating the resources required for computation, also known as performance modeling, plays a central role in classical high-performance computing.
Elaborate techniques have been developed to understand application scaling characteristics, their portability across architectures, and resource consumption. 
These models consider some resources as a measure of cost, such as time, energy, number of instructions, and more recently, data movement. 
In this section, we first reflect on the mature field of classical computational resource estimation, and then leverage this perspective to present a framework for quantum resource estimation.

\begin{figure}[b]
	\centering
	\includegraphics[width=1.0\textwidth]{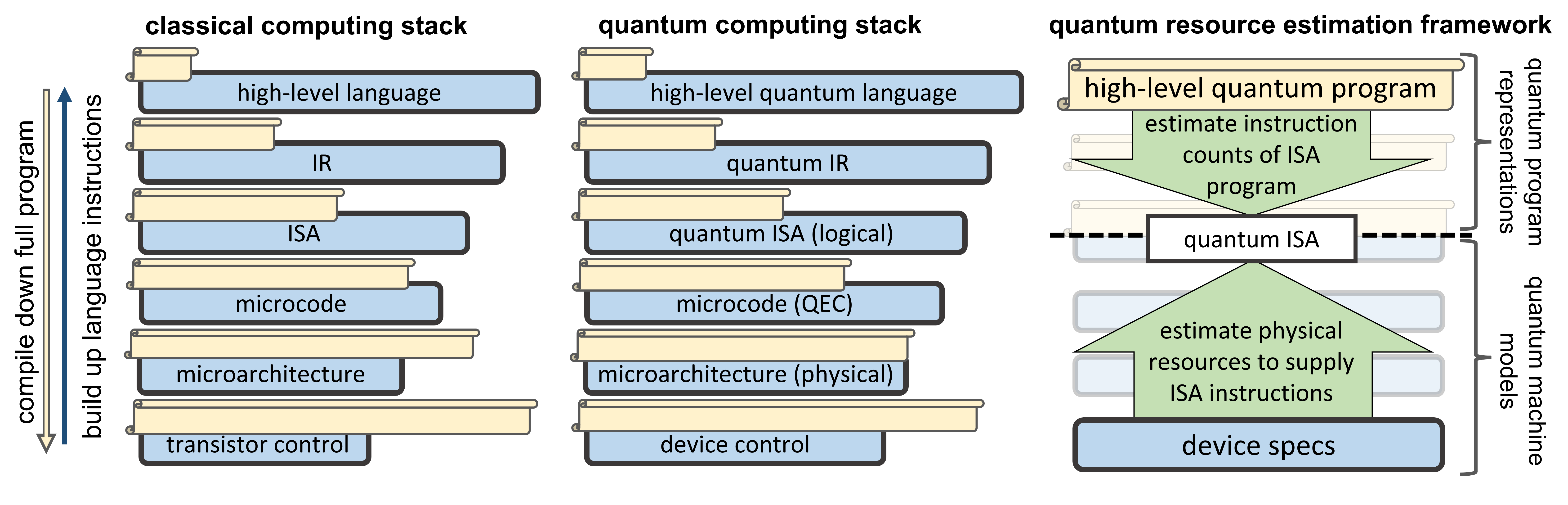}
  \caption{
    The stacks for classical (left) and quantum (center) computing.
	At the top of the stack the program is expressed in a natural, high-level language such as C/C++ (classical) or Q\# (quantum).
	This program is re-expressed as different program representations (manilla scrolls) in a sequence of languages (blue boxes) which are represented here as layers, with each layer being a lower-level language than that above, down to the physical signals controlling the device.
	A complementary viewpoint is that the instruction set available to the language at one layer can be combined to form a more elaborate instruction set for the layer above.
	The scrolls are shortest at the top, symbolizing that program representations are simplest when expressed in higher-level languages, while the blue boxes are smallest at the base of the stack, symbolizing that instruction sets are simplest at the lower levels.
	(right)
	Our quantum resource modeling framework is a modular representation of the layers of the quantum computing stack, which collects together the upper layers
	as \emph{quantum program representations}, and the lower layers as \emph{quantum machine models}, with the quantum ISA connecting both. 
	}
	\label{fig:program-flow}
\end{figure}

In the classical computing stack, the program is initially expressed in a high-level language at the top and is sequentially re-expressed in more and more explicit representations down the stack; see the left column of \fig{program-flow}. 
We can consider each layer in the stack as having a language with a specific \textit{instruction set} of allowed instructions, which can be broken down in terms of simpler instructions in the layer below, or combined together to form more abstract instructions for the layer above.
For example the program could be expressed in C++, which is compiled into an intermediate representation (IR) for optimizations, e.g., LLVM's static single assignment form (SSA) intermediate representation~\cite{LLVM}. 
The IR is then lowered to assembly code in a specific instruction-set architecture (ISA) that is the interface to the classical processor. 
Such assembly code could consist of x86~\cite{x86}, ARM~\cite{arm}, or RISCV~\cite{riscv} instruction sets. 
Some complex instruction set computers such as Intel's Haswell translate the x86 code into a microcode representation for execution by the arithmetic logical units (ALUs).
The microarchitecture then executes those micro-instructions and eventually translates them to electrical signals to be implemented by transistors. 
For some processors, microcode and microarchitecture are combined, but we illustrate them in separate layers here to draw parallels with the quantum stack. 

The goal of resource estimation is to model the stack in order to estimate some physical resources such as the run time
required to implement a high-level program. 
One approach could be to consider a chain of re-expressions of the program from the very top of the stack all the way to the bottom and then read off the physical resources needed.
Near the top of the stack this is an appealing approach as it is very natural to think of a program being compiled down into a sequence of lower-level representations. 
Towards the bottom of the stack however this approach becomes very difficult because of the intricacies of the microarchitecture, runtime dependencies of the program execution, and the overall length of the representation itself.  
However, the instruction sets within which the program is expressed typically become simpler for lower layers, which can be exploited to model resources more conveniently.
The strategy is then to choose a layer towards the middle of the stack, and model the layers above by re-expressing the high-level program down to the instruction set at this layer, and estimate the physical resources required to implement those instructions by building them up from the more basic instructions in the layers below.

There is of course a choice about where to draw the line between the top and the bottom in this view of the stack, and the best choice can depend on the precise goals of the model. 
For most purposes, the ISA layer has emerged as the natural interface because it provides an implementation-independent interface that can be used to ground modeling efforts. 
ISAs such as x86, ARM and RISCV represent program executions across a wide variety of classical processors. 
Performance modeling frameworks can leverage this powerful abstraction to model the resources of the full stack by first establishing the requirements of the program when expressed as instructions the ISA level, and then identifying the resources required at the device level to implement the ISA-level program. 

The quantum computing stack is less established and abstractions are still in their definition phase. 
Some aspects are similar, for example while a quantum program includes instructions which are executed as quantum operations, the quantum program itself is just a sequence of instructions similar to a classical program.
However, in stark contrast with the classical stack, which is built upon extremely reliable transistors, the quantum stack must account for the unavoidable fact that all forms of physical qubit implementations are inherently noisy~\cite{Preskill2018NISQ}. 
To execute a quantum application successfully, quantum error correction (QEC)~\cite{gottesman1998} must be used to build logical qubits that can be used to store and manipulate quantum information better than raw physical qubits. 
This QEC capability is central to scalable quantum computers, but the costs are formidable, 
often multiplying the number of qubits needed by a factor of thousands, and runtimes by a factor of hundreds.
Therefore, the specific QEC approach used impacts design decisions across the quantum computing stack, both at the hardware layers which must be capable of implementing it, and at the software layers which must compile to logical operations compatible with it. 

Prior works that address the design of a flexible quantum stack and resource estimation span vastly different viewpoints. 
For example, prior quantum architecture and compilation proposals \cite{ibmstack, triq, shicompilation, bertels} develop stacks for near-term small quantum computers, but do not incorporate QEC. 
These works either ignore noise~\cite{bertels}, or reduce its impact with error mitigation techniques~\cite{ibmstack} which unfortunately do not scale to larger systems~\cite{quek2022}.
On the other hand, works comparing resources for multiple QEC codes and logical operations, such as \cite{beverland2021,Chamberland2022b,Chamberland2019,Bombin2021,Litinski2019b}, typically address lower layers of the stack without incorporating the software layers.
There is a growing number of informative end-to-end resource analyses, but typically these single out very specific algorithms and hardware and make very different assumptions across the stack, making direct comparisons of different approaches challenging~\cite{Fowler2012,Gidney2021,gouzien2021,Chamberland2020, rotteler17, beauregard03, vanmeter10,lee22, suchara13}. 

\subsection{A framework for quantum resource estimation}

We present a quantum resource estimation framework that aims to capture the essential aspects of a full-stack quantum computer with the flexibility to allow direct and easy comparisons across a range of algorithms, software techniques, QEC codes, hardware specifications and other aspects.
To form the skeleton of our framework, we propose the stack shown in \fig{program-flow} (center) for large-scale quantum computers. 
This framework builds upon the previous work mentioned above, but seeks to clearly delineate the different layers, clarify the functions of each layer and define how they inter-operate to attain the computation capabilities required for large-scale applications. 
We find that it is possible to draw broad parallels between the classical and quantum computing stacks, allowing us to leverage established resource modeling strategies from classical performance modeling.

In what follows we describe the layers in this stack, briefly stating the main characteristics of each layer, and listing some known examples from the quantum computing literature which naturally fit into it.
As we will see, the layers are modular, allowing them to be designed either collectively or independently, provided that the interfaces between layers remain consistent. 
Consistency of the interfaces is ensured by providing explicit maps: compilation algorithms for program representations down to the ISA from above, and explicit constructions of instruction sets from those below up to the ISA from below.
These maps are themselves a crucial part of the stack and the modeling framework.
We start at the top of the stack, where the layers are conceptually similar to the classical case. 

\textbf{High-level quantum language.}---
This layer hosts the explicit representation of an algorithm as a program expressed in a high-level programming language.
The language should enable the expression of any quantum algorithm. 
At this level of abstraction, users need not know the requirements of quantum error correction, and may assume their program and its operations are fault tolerant.
Languages ideally include support for loops, functions, rich type systems, debugging and other functions that are common in classical computing in order to allow quantum applications to be developed and tested efficiently. 
Similarly, supporting hybrid quantum-classical computation which allow applications to use phases of classical and quantum operations seamlessly is valuable for practical applications. 
Examples of languages include Q\#~\cite{svore2018}, Quipper~\cite{green2013}, Scaffold~\cite{abhari2012}, QWire~\cite{paykin2017}, Quil~\cite{quil} and others. 

\textbf{Quantum IR.}---
To execute a high-level quantum program, that program's instructions need to be \emph{compiled} and expressed in terms of quantum ISA instructions.
It is appealing to support executions of programs expressed in a diverse set of high-level languages across a range of hardware back ends that may have different quantum ISAs.
In analogy with classical computing, we propose the use of a quantum intermediate representation (quantum IR) that expresses high-level operations in a language-agnostic and ISA-agnostic manner~\cite{mccaskey2021,Ittah2022,qir}.
A compiler then includes three components --- the front end, the optimization layer and the back end. 
The front end component translates high-level language instructions to the quantum IR. 
The back end component translates the quantum IR into the quantum ISA operations. 
The middle optimization layer performs transformations on the quantum IR to reduce resource requirements~\cite{Ittah2022}.
To support a new quantum language or quantum ISA, we simply need to add the corresponding front end or back end components, rather than reinventing the whole compiler. 
An example of a quantum IR is the aptly named QIR~\cite{qir}, which includes support for expressing logical operations, measurements and other quantum constructs that are required for applications. 

The parallels of these two upper layers with the classical stack can be leveraged. 
For instance, later we consider high-level quantum programs expressed in Rust, and the classical IR LLVM has been applied to quantum compilation tasks Ref.~\cite{javadiabhari2014,Meuli2020}.
Next we address the layers at the bottom of the stack and describe how their instruction sets build up to form the instruction set of the logical qubits at the quantum ISA layer (which the quantum IR compiles down to).

\textbf{Device control.}---
There is a wide range of hardware approaches to build the quantum device at the base of the stack, including those based on photonic~\cite{knill2001,carolan2015,bombin2021a}, trapped atom~\cite{Jaksch1999,bluvstein2022}, trapped ion~\cite{Cirac1995,pino2021}, Majorana~\cite{kitaev2001,sarma2015,Karzig2017}, spin~\cite{kane1998,Hanson2007,Jnane2022}, superconducting transmon~\cite{transmonSchoelkopf,schreier2008,kjaergaard2020,arute2019,hong2020,Steffen2011}, and electro-acoustic~\cite{Guillaud2019,Chamberland2022a} platforms.  
Quantum devices themselves are not sufficient for computation, we require classical computing resources to bring up, calibrate and control the devices as well as the ability to transmit and process information from quantum measurements.
The nature of how qubits are stored within the device, the range of control available, and the nature of that control are all very device-dependent and form the instruction set at the device level -- the constraints on ion traps will be different than the constraints on transmon qubits, for example, simply because their physical realizations are fundamentally different.
It is at this level that resources are typically most forthright - one may consider run time, number of qubits, control bandwidth, device size and other cost metrics.

\textbf{Microarchitecture (physical).}---
In this layer, low-level details of the qubit device design are abstracted away, leading to a viewpoint consisting of a set of abstract physical qubits, along with a discrete set of physical instructions such as single-qubit measurements and CNOT gates. 
These instructions can be characterized by simple properties such as their action on the physical qubits, how long they take to apply, and their probability of failure (their so-called \emph{error rate}).
This abstraction allows microarchitectural components to be reused across a variety of qubit device implementations, while flexibly supporting the needs of different QEC schemes.
Each of the operations in the physical instruction set is built from operations available at the device control level (for example, in ion traps a CNOT may be implemented by a sequence of M\o{}lmer-S\o{}rensen~\cite{Molmer1999} gates, while in spin qubits it may be implemented by a sequence of exchange interactions~\cite{Hanson2007}).
In planar device platforms like superconducting qubits or Majorana qubits, the ability to support such device-level operations can lead to connectivity restrictions such as 2D nearest-neighbor connectivity at this level. 

\textbf{Microcode (QEC).}---
The microcode instruction set consists of the primitives required to implement a QEC strategy, which uses a QEC code to form reliable logical qubits using noisy physical qubits. 
These logical qubits support logical operations (which form the instruction set for the quantum ISA level above) that have better error rates than raw physical operations. 
For example, if the QEC strategy uses the surface code~\cite{Kitaev2003,Bravyi1998}, the primitives which form the instruction set will include \textit{syndrome measurement circuits} expressed in the physical instruction set of the microcode layer below.
For the surface code, syndrome measurements provide the information required to diagnose and correct faults at the physical level by measuring stabilizers, but can also induce logical operations by deforming the code~\cite{Litinski2019,Horsman2012}.

\textbf{Quantum ISA (logical).}---
This layer forms the interface between the software and hardware layers. 
It abstracts the details of how QEC is implemented in the layer below, retaining only a set of fault-tolerant logical operations as its instruction set.
It is crucial that the instruction set is \textit{universal}, meaning it is complete for quantum computing.
In this formulation of the stack, it is the map from the microcode level to the quantum ISA level that actually implements the error correction, using a classical algorithm known as a \textit{decoder} to identify and correct faults while implementing specific logical operations.
Clearly, the instructions in this layer must be chosen such that they can implemented by the underlying hardware; that is, when choosing an ISA, it is important to consider the constraints of the system architecture such as connectivity of logical qubit representations and ability to construct the logical operations with known sequence of micro-operations and physical qubit control operations. 
At the same time, the ISA operations should be chosen such that the software layers above can efficiently express common application operations using it. 
The planar quantum ISA is an example we will discuss later which arises from the logical operations of two different QEC approaches, one based on surface codes and the other on Hastings-Haah codes.
For instance, the planar quantum ISA instruction set includes single-qubit initializations, single-qubit measurements and multi-qubit measurements, which are all examples of Clifford operations.
Clifford operations by themselves are not universal~\cite{GottesmanKnill}, but when combined with a fault-tolerant T state initialization, the resulting set becomes universal.

\textbf{Estimating resources.}---
Our goal here has been to define the layers of the quantum stack at a sufficiently high-level of abstraction, so that a broad range of known and yet-to-be discovered approaches can be readily incorporated. 
This forms the basis of a \emph{modular} framework for quantum resource modeling, in which layers can be considered together or independently.
We find it convenient to -- similar to the classical case -- split the quantum stack in two at the quantum ISA and analyze the upper layers from the top down in terms of programs, and the lower layers from bottom up in terms of instruction sets; see \fig{program-flow} (right).
To use the framework to make explicit estimates, one models a consistent instantiation of each layer and each of the maps between layers with an explicit example.
Later, we will specify examples of these layers and maps in more detail to answer questions regarding system requirements for scaling. 

This framework enables the estimation of resources (such as the number of qubits, the run time and the power consumption) which would be required to implement a given quantum algorithm using a given qubit technology and with a fixed set of architectural choices.
Within the framework, architectural choices can be viewed as explicit specifications for each layer and map in the hardware and software parts of the quantum stack. 
There is no single answer to questions on architectural design trade-offs, with different choices resulting in different estimation results which can be explored to make informed decisions as researchers and engineers work to scale up a complete system.
One can, for example, trade off more qubits against shorter run times, or trade off faster qubit gate operations against lower fidelities.
One can also use the framework much more broadly, to compare new error correction proposals, to evaluate the potential for quantum advantage for new or optimized quantum algorithms, and to motivate future research directions by identifying bottlenecks in the stack which generate large resource contributions.

The modular nature of the framework allows for future modifications and extensions.
For some purposes it may be desirable to combine some layers together, and to split other layers into sub-layers.
In this work our focus is on resource estimates for the quantum accelerated parts of the computation, however more layers can be added to model the hybrid aspects of the computation. 

\subsection{Fault-tolerant design considerations}

Before discussing specific implementations of the quantum resource estimation framework, we address several significant design considerations that arise from the need for QEC.
Typically, a QEC approach is centered around a QEC code family which stores the logical qubits.
A wide range of QEC codes have been discovered over the last three decades with wide ranging properties, including small codes such as the Shor~\cite{shor1995} and Steane~\cite{steane1996} codes, 2D local codes such as surface codes~\cite{Kitaev2003,Bravyi1998}, color codes~\cite{bombin2006}, Hastings-Haah codes~\cite{Hastings2021} and Bacon-Shor codes~\cite{bacon2006,shor1996}, and more exotic families of positive-rate low-density parity check codes~\cite{Tillich2014, panteleev2022,guth2014,leverrier2015}. The choice of QEC code has impact up and down the stack since different codes can place different implementation demands on the physical instructions provided by the microarchitecture level below, while also exposing different logical operations to the quantum ISA level above.

Here we focus on the requirements that QEC imposes on the physical operations at the microarchitecture level of the quantum stack, corresponding to the setting known as \textit{circuit-level} analysis in the QEC literature.
It is common in this setting to assume QEC is implemented with Clifford operations (state preparation, Clifford unitary gates, and Pauli measurement) for single qubits and across pairs of qubits coupled by a specified \textit{connectivity}, and that those operations fail independently with a uniform error rate (see \app{noise}).
An important metric for a QEC approach in this setting is its \textit{threshold}, which specifies the maximum error rate that it can tolerate.
In what follows, we leverage known threshold results for a range of QEC schemes to extract a number of key design parameters pertaining to error rates, parallelism and connectivity.

The theoretical upper limit for any threshold is not known, but the highest threshold discovered to date is about 3\% using the scheme described in Ref.~\cite{knill2005}, and a range of schemes~\cite{Wang2011, beverland2016, tremblay2022, Paetznick2022} have been found with thresholds approaching 1\%.
To avoid prohibitive QEC overheads, error rates at least an order of magnitude smaller than the threshold are much preferred. 
For this reason, we require physical error rates on Clifford operations below 0.1\%.

The aforementioned schemes obtain high threshold values only in the setting where operations can be applied in parallel, which may pose a significant hardware challenge for some platforms such as trapped ions~\cite{Wang2017}. 
However, the necessity of parallel operations has long been believed necessary for any QEC scheme to exhibit a non-zero threshold~\cite{steane1998,gottesman2010}.
We therefore require that physical operations can be applied in parallel.

Next we consider connectivity. 
Probably the most practically relevant classes of connectivity correspond to arrays of qubits with neighboring connections in one or two dimensions, which we refer to as \textit{1D} and \textit{2D connectivity} respectively. 
For connectivity with any fixed dimension, the thresholds of many of the more exotic code families are believed to drop to zero~\cite{Bravyi2010,delfosse2021}.
With 2D connectivity the thresholds of some QEC schemes based on concatenating small codes are reduced~\cite{svore2005}.
Fortunately, many 2D local codes (including the surface code~\cite{Wang2011}, the Hastings-Haah code~\cite{Paetznick2022} and the color code~\cite{beverland2016}) are naturally implementable with 2D connectivity and achieve high thresholds approaching 1\% in this setting despite the connectivity restrictions.
Further restricting to 1D connectivity leaves very few code families known to have finite thresholds, and those that do are around 0.01\% and below~\cite{jones2018,szkopek2006,stephens2009}.
Moreover, 1D connectivity is vulnerable to hardware defects that some qubits permanently unusable.
To avoid such potentially fatal conditions, we require that qubits are well connected, for example with 2D connectivity but not 1D connectivity.

There are a number of important subtleties that we have ignored so far.
First, the simple uniform discrete error model is not a true refection of most systems. 
Including some realistic aspects of noise, such as asymmetries in the probabilities of different error channels~\cite{Brown2020,Chamberland2022a} and richer information from noisy measurements~\cite{pattison2021} can improve error correction performance.
On the other hand, including other realistic aspects, such as correlated errors~\cite{klesse2005} and permanent hardware defects which must be worked around to implement QEC~\cite{strikis2021,Stace2009} can worsen error correction performance.
To partially allow for these subtleties, we require that the \emph{worst} error rates for Clifford operations that are employed by the QEC approach are below 0.1\%.
Next, we have required the qubits are \textit{well connected} to enable practical QEC, and make it clear that 2D connectivity falls into this class, while 1D connectivity does not.
However, we leave the formulation of a more complete definition which applies to more general connectives to future work.

In summary, we require physical qubits to be well connected such as having 2D connectivity, with the ability to perform parallel operations with error rates below one part in a thousand. 
This holds for all examples we consider in this paper, and forms the basis of one of our criteria for scale.

\section{Estimating resources for three quantum applications}
\label{sec:examples}

In this section we introduce an implementation of the quantum resource modeling framework, hereafter referred to as `the tool', which is publicly available in Microsoft Azure Quantum as the first version of the Azure Quantum Resource Estimator~\cite{toolLink}.
Our primary goal in this work is to identify architecture features which will be crucial to achieve practical quantum computing advantage.
To do so, we estimate the resources for select applications with the potential for practical quantum advantage (\sec{application-examples}), using qubit parameters which are relevant for a number of prominent qubit technologies.
We fix a set of consistent architectural options for the quantum stack based on established approaches from the literature.

\begin{figure}[b]
	\centering
	\includegraphics[width=1.0\textwidth]{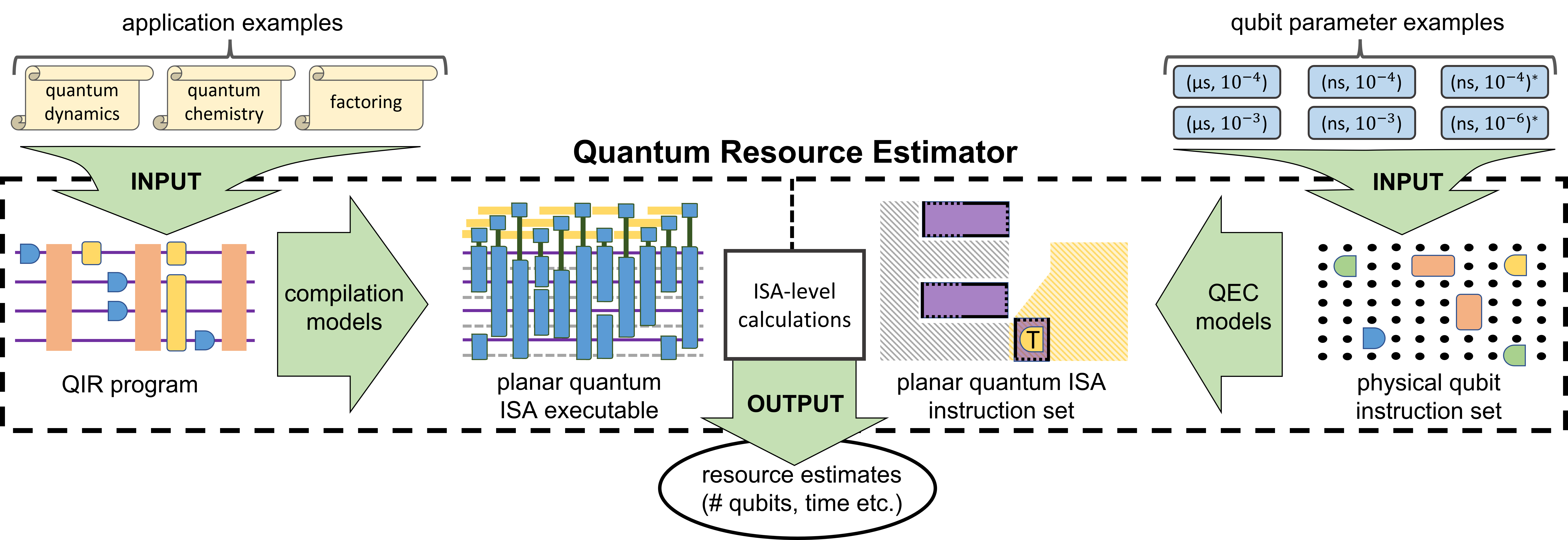}
	\caption{
    A sketch of how the tool estimates resources for the three example applications and six qubit parameter examples we consider.
    Note the stack is represented left-to-right rather than top-to-bottom.
    Applications, addressed by a high-level quantum program, are first translated explicitly to a QIR program which is input into the tool, where its compilation down to an ISA-level executable is modeled.
    All our examples flow to the same planar quantum ISA, which has an instruction set consisting of logical surface code operations.
    On the hardware side, we input physical qubit parameters to the tool including the physical qubit instruction set and the times and error rates of those instructions. 
    The tool then models how these noisy qubits are used to build up protected logical qubits and to fault-tolerantly implement the planar quantum ISA instructions using configurable QEC models.
    The tool outputs resource estimates such as the number of physical qubits and time required to run the application.
    }
	\label{fig:quantum-resource-estimator}
\end{figure}

Following our framework, estimates obtained using the tool can be viewed as consisting of two broad components as shown in \fig{quantum-resource-estimator}.
The first component (\sec{algorithms-and-software}) involves an explicit front end compilation from a high-level language to a quantum IR, which is fed to the tool.
The tool then models the back end compilation down to a quantum ISA, yielding ISA-level resource estimates including the number of logical qubits and the number of logical time steps.
In the second component (\sec{hardware-and-QEC}), specific qubit parameters are fed to the tool which performs a bottom-up estimation of the hardware resources required to implement ISA instructions by modeling error correction with specific QEC codes.
In \sec{putting-pieces-together} we combine these components to obtain concrete resource estimates. 

In the appendices we provide a more detailed and self-contained description of the material in this section, including our modeling assumptions.
Our examples are available as samples at \cite{samplesLink}.

\subsection{Quantum applications}
\label{sec:application-examples}

The first of our three example applications is a \emph{quantum dynamics} simulation that is one of the smallest scientifically interesting problems that is out of reach for classical computation. 
Specifically, we consider a 2D transverse-field Ising model with 100 quantum spins, propagated for ten time steps using a fourth-order Trotter algorithm as described in Ref.~\cite{Pearson2020}. 
This is among the simplest models to exhibit a quantum phase transition. Its dynamics are representative of strongly entangled quantum systems and as such is believed to be classically intractable~\cite{bauer2020quantum}.

Our second example is a \emph{quantum chemistry} application to calculate the energy of a ruthenium-based catalyst for carbon fixation which could have implications for reversing the effects of global warming.
More specifically, to estimate the energy of Complex~XVIII in Ref.~\cite{vonBurg2021} to chemical accuracy of 1~mHa using the so-called `double-factorized qubitization' algorithm described in Ref.~\cite{vonBurg2021}. 

Our third application is \emph{factoring}, specifically to identify the pair of prime factors of a $2048$-bit integer.
Solving this problem would form the basis of an attack on widely used RSA-based encryption schemes.
In our resource estimation model, we use a high-level program to implement Shor's factoring algorithm based on that described in Ref.~\cite{Gidney2021}.

To achieve the desired solution accuracy for each application, we also set a minimum probability that the overall algorithm succeeds, which we call the \textit{algorithm execution accuracy}.
For the quantum dynamics and quantum chemistry examples we assume that algorithm execution accuracies of $0.999$ and $0.99$ will be required.
However factoring admits a lower algorithm execution accuracy since it is easy to check if a proposed solution is correct (by multiplying the purported prime factors to identify if they recover the original integer), and so we take it to be $2/3$ in this case.

\subsection{Requirements at the quantum ISA level}
\label{sec:algorithms-and-software}

Here we describe how we obtain ISA-level resource estimates needed to target our three example applications.
To identify the physical-level costs of implementing an ISA-level executable, we will not only need to know parameters capturing the size of the program, but also the required quality of the ISA operations needed to achieve the desired algorithm execution accuracy.
This is because different quality requirements will result in different error correction solutions being preferred.

As a starting point, we assume explicit high-level implementations of the quantum-accelerated components of each of the three target applications, using Rust for quantum chemistry and factoring and Q\# for quantum dynamics\footnote{These implementations are available as samples at \cite{samplesLink}}. 
First we use an explicit front end compilation to a specific quantum IR, namely QIR~\cite{qir} using the Q\# compiler or the Rust compiler based on the application.
The tool then models the back end stage of compilation which translates QIR into operations supported by the quantum ISA.

In this work we use the \textit{planar quantum ISA}, which is based on the logical operations of the most well-established QEC code, namely the surface code~\cite{Kitaev2003,Bravyi1998}, and also applies to the recently developed Hastings-Haah codes~\cite{Hastings2021,Paetznick2022}. 
Logical qubits are stored in patches laid out in a 2D plane, along with ancilla regions.
We sketch some key features of the planar quantum ISA instructions in \fig{quantum-resource-estimator}, and provide a full specification in \fig{logical-operations} in \app{logical-operations-details}.
Only Clifford operations involving patches (purple) that are connected by ancilla regions (grey) are permitted, and two such operations can only be performed simultaneously only if the ancilla regions they require are disjoint. 
Non-Clifford T states are produced in dedicated T factories (yellow).

There are two primary conceptual challenges that need to be overcome to implement the back end compilation phase.
Firstly, the QIR instruction set contains fine-angle rotation unitaries, which must be approximated with sufficient accuracy by sequences of operations from the planar quantum ISA,
a process known as \textit{synthesis}.
Secondly, one requires methods to layout qubits and perform communication in order to map from the QIR, which has no connectivity restrictions, to the planar quantum ISA, which has geometric constraints. 
Standard layout and communication methods can be applied to address this second aspect of compilation, including SWAP-based schemes where qubits are moved by swapping neighbors, as well as teleportation-based schemes where qubits are moved using quantum teleportation. 
In \app{cat-state}, we provide an alternative back end compilation scheme which we refer to as Parallel Synthesis Sequential Pauli Computation (PSSPC), which is based on a combination of favorable features from existing approaches~\cite{Litinski2019,Chamberland2021,Beverland2022,Kliuchnikov2022}.
The tool first takes a trace of the QIR program and then uses accurate formulas to calculate the planar quantum ISA resource requirements that would result from back end compiling with PSSPC.

\tab{application-examples} presents the ISA-level resource requirements of our three example applications, which are seen to vary significantly across the applications.
Our quantum dynamics example, selected as being among the smallest applications of scientific interest, requires only 200 logical qubits after mapping to the planar quantum ISA. 
However, even this application requires logical qubits with error rates that are below $10^{-11}$, which puts it beyond the capabilities of current noisy quantum hardware. 
Applications like chemistry and factoring require several thousand logical qubits, with each logical qubit having error rates in the range of $10^{-15}$ to $10^{-18}$ and T states with error rates in the range of $10^{-12}$ to $10^{-15}$. 
These applications require mature QEC implementations and large-scale, reliable quantum hardware which we discuss in the next subsection.

\begin{table}[h]
	\begin{tabular}{|c||c||c|c|c||c|c|}
		\hline
		  \multirow{2}{*}{application} & algorithm execution &   \multicolumn{3}{|c||}{quantum executable parameters} & \multicolumn{2}{|c|}{quality requirements}   \\
		\cline{3-7}
		 & accuracy $1-\epsilon$ &   $\Q$ & $\TimeLogicalFast$ & $\R$   & max $\Plog$   & max $\PlogDis$   \\
		\hline
		\hline
		quantum dynamics & $0.999$ &  230 & $1.5 \cdot 10^{5}$  & $2.4 \cdot 10^{6}$  & $9.7\cdot 10^{-12}$ & $1.4 \cdot 10^{-10}$  \\
		\hline
		quantum chemistry & $0.99$ & 2740 & $4.1 \cdot 10^{11}$ & $5.4 \cdot 10^{11}$  & $3.0\cdot 10^{-17}$ & $6.1\cdot 10^{-15}$  \\
		\hline 
		factoring & $0.667$ & $25481$ & $1.2\cdot 10^{10}$ & $1.5\cdot 10^{10}$  & $3.5\cdot 10^{-16}$ & $7.4\cdot 10^{-12}$  \\
		\hline
	\end{tabular}
	\caption{
		Estimates of the ISA-level requirements to implement our applications.
		The application is compiled down from a high-level application program to a program expressed in the quantum ISA, with parameters as shown. These requirements include the number of logical qubits $\Q$, the minimum number of logical time steps (also called logical time steps) $\TimeLogicalFast$ and the number of T states $\R$ consumed by the program. To ensure that the algorithm fails with probability at most $\epsilon$, the tool also estimates the maximum allowed error rate for each logical qubit $\Plog$ and the maximum allowed error rate for the distilled T states $\PlogDis$. 
        }
	\label{tab:application-examples}
\end{table}

\subsection{Requirements at the device level}
\label{sec:hardware-and-QEC}

To understand the high-level impact of physical qubit parameters on application resource estimates, we configure the tool to model the underlying qubit technology relatively abstractly in terms of operation times and error rates, and consider parameter configurations which are relevant for a range of prominent hardware approaches.
These features are captured at the microarchitecture level of the quantum stack.
Along with the operation time and error rate, the tool takes as input one of two physical operation sets: \textit{gate-based instructions} (including CNOTs and single-qubit measurements) relevant for technologies such as superconducting qubits, trapped ion qubits etc., and \textit{Majorana instructions} (including non-destructive two-qubit Pauli measurements) which are relevant for Majorana qubits~\cite{Karzig2017,Plugge2016}. 
Both of these instruction sets have 2D connectivity and allow the parallel application of disjoint operations.

\begin{table}[h]
	\begin{tabular}{|c||c|c|c|c|}
		\hline
		\textbf{qubit parameter}	& \multicolumn{2}{|c|}{operation times}		& \multicolumn{2}{|c|}{error rates} \\
		\cline{2-5}
		\textbf{examples}& 	gate	&	measurement		& Clifford &  non-Clifford \\
		\hline
		\hline
		\textbf{$(\boldsymbol{\mu}$s,~$\mathbf{10^{-3}}$) qubit}	 &	100 $\mu$s &	100 $\mu$s		& $10^{-3}$ & $10^{-6}$\\
		\hline
		\textbf{$(\boldsymbol{\mu}$s,~$\mathbf{10^{-4}}$) qubit}	 &	100 $\mu$s &	100 $\mu$s		&  $10^{-4}$ & $10^{-6}$ \\
		\hline
		\textbf{(ns,~$\mathbf{10^{-3}}$) qubit}		 &	50 ns &	100 ns		& $10^{-3}$ &  $10^{-3}$\\
 		\hline
		\textbf{(ns,~$\mathbf{10^{-4}}$) qubit}		 &	50 ns &	100 ns		& $10^{-4}$ &  $10^{-4}$\\
 		\hline 		
		\textbf{(ns,~$\mathbf{10^{-4}}$) Majorana qubit}	 &	100 ns &	100 ns		&  $10^{-4}$ & 0.05\\ 
		\hline
		\textbf{(ns,~$\mathbf{10^{-6}}$) Majorana qubit}	 &	100 ns &	100 ns		&  $10^{-6}$ & 0.01\\
		\hline
	\end{tabular}
	\caption{
	We consider a set of examples of qubit parameters which represent various regimes of interest.
	Each example is labelled with the units of its operation times (either $\mu$s or ns), and the limiting error rate of its Clifford operations.
	The parameter values and instruction set chosen for each example is informed by published proposals for various hardware approaches. 
	We assume that the first four examples have gate-based instruction sets, while the last two examples have Majorana instruction sets.
    }
	\label{tab:qubit-type}
\end{table}

As listed in \tab{qubit-type}, we consider six qubit parameter examples, each labelled with the unit of its operation time and error rate.
In \fig{quantum-resource-estimator}, we use an asterisk to differentiate the two examples which use Majorana-based instructions rather than gate-based instructions. 
With operation times in the microsecond range, the ($\mu$s,~$10^{-3}$) and the ($\mu$s,~$10^{-4}$) qubits are relevant for trapped ions~\cite{Cirac1995}, while the (ns,~$10^{-3}$) and (ns,~$10^{-4}$) qubits are more relevant for superconducting transmon qubits~\cite{transmonSchoelkopf} or spin qubits~\cite{kane1998}. 
We also include the (ns,~$10^{-4}$) and (ns,~$10^{-6}$) Majorana qubit examples, to account for future topological qubits based on Majorana zero modes~\cite{Karzig2017}.
These latter qubits are also expected to have operations in the nanosecond regime, and owing to topological protection in the hardware, they also have the potential for higher fidelities at the physical level.
In \app{noise} we provide more explanation of these qubit parameter examples and provide the full specifications of the aforementioned instruction sets in \fig{physical-operations}.

In this work, we assume that logical qubits are encoded in patches (purple in the ISA instructions of \fig{quantum-resource-estimator}) of either the surface code~\cite{Kitaev2003,Bravyi1998} or the recently discovered Hastings-Haah code~\cite{Hastings2021}.
We make this restriction primarily because these codes have high fault-tolerance thresholds, have relatively well-understood logical operations, and can be implemented using physical qubits with 2D connectivity.  
We model QEC to relate properties at the ISA (logical) level to properties at the microarchitecture (physical) level.
For both the surface code and Hastings-Haah code, error suppression can be tuned by scaling the patch's code distance $d$.
For fixed qubit parameters, increasing the distance costs quadratically more physical qubits $n(d)$ and has a linearly longer logical time step $\tau(d)$, but reduces the probability $\Plog(d)$ of a logical qubit failing during a logical time step exponentially.
A number of fault-tolerant logical Clifford operations have essentially no time cost in these codes, including state initialization, readout and Pauli gates. 
On the other hand, logical qubit movement and multi-qubit Pauli measurements are implemented via lattice surgery with the use of ancilla regions (grey in \fig{quantum-resource-estimator}) and require a logical time step. 
For counting resources, it is convenient to partition the physical qubits into tiles, where each tile contains $n(d)$ physical qubits, and can either be used to encode a logical qubit in a code patch, or can form part of an ancilla region.
The total number of physical qubits used for code patches and ancilla regions is then $\Q \cdot n(d)$, where $\Q$ is the number of tiles (which we also refer to as the number of logical qubits). 
See \app{logical-operations-details} for details.

For the ISA-level instruction set to be universal, we also require non-Clifford operations, such as a logical T gate.
This is achieved via the generation of T states (also known as magic states) using T factories (yellow in \fig{quantum-resource-estimator}), which typically requires many more resources than other quantum ISA operations. 
These T factories typically involve a sequence of rounds of distillation, where each round takes in many noisy T states encoded in a smaller distance code, processes them using a distillation unit, and outputs fewer less noisy T states encoded in a larger distance code, with the number of rounds, distillation units, and distances all being parameters which can be varied. 
There are therefore many options when selecting a T factory. 
We use the symbol $\mathcal{D}$ to label a particular T factory choice which requires $n(\mathcal{D})$ physical qubits, and takes a time $\tau(\mathcal{D})$ to produce a T state with error rate $\PlogDis(\mathcal{D})$. 
We assume T factories that are space and time efficient implementations of the well-known 15-to-1 distillation circuit; see \app{distillation-overhead} for details.  

For each application, the code distance $d$ and T factory $\mathcal{D}$ must be chosen to ensure that the ISA requirements from \tab{application-examples} are satisfied, namely that $\Plog(d) \leq \text{max}~P$ and $\PlogDis(\mathcal{D}) \leq  \text{max}~\PlogDis$, but with minimal cost. 
As an example, \tab{hardware-examples} shows the resource costs to satisfy the ISA-level requirements for the factoring application across different physical qubit parameters. 
We can see that the number of physical qubits and the duration of a logical time step reduce as physical error rates improve. 
For this example application, improving physical error rates from $10^{-3}$ to $10^{-4}$ can reduce the number of qubits by a factor of four, and can halve the logical time step. 
Comparing qubits with microsecond and nanosecond operating times, we see the impact of physical operation time at the ISA level  --- logical time steps are several orders of magnitude higher than physical operating times, underscoring the importance of fast physical operations. 
T factories incur significant physical overheads, requiring several thousand physical qubits and only producing new T states once every 10 to 15 logical time steps, motivating the need for multiple parallel T factories to meet application T state demand.

\begin{table}[h]
\begin{tabular}{|c||c||c|c|c||c|c|}
\hline
& & \multicolumn{3}{c||}{logical qubit parameters}                    & \multicolumn{2}{c|}{T factory parameters}                                              \\ 
qubit parameter  & QEC code    & \multicolumn{3}{c||}{for $\Plog(d) \leq 3.5\cdot{10^{-16}}$}                    & \multicolumn{2}{c|}{for $\PlogDis(\mathcal{D}) \leq 7.4\cdot{10^{-12}}$} \\ 
\cline{3-7}
 examples & selected & distance  & \# qubits& logical time step     &\# qubits   & duration \\ 
 &  & $d$  & $n(d)$ & $\tau(d)$    & $n(\mathcal{D})$   & $\tau(\mathcal{D})$ \\
\hline
\hline
($\mu$s,~$10^{-3}$) qubit     & surface & 27 & 1458   & 16 ms     & 17640                            & 163 ms                          \\ \hline
($\mu$s,~$10^{-4}$) qubit      & surface & 13 & 338    & 7 ms  & 4840                            & 85 ms                           \\ \hline
(ns,~$10^{-3}$) qubit     & surface & 27 & 1458   & 10 $\mu$s    & 33320                           & 128 $\mu$s                          \\ \hline
(ns,~$10^{-4}$) qubit    & surface  & 13 & 338    & 5 $\mu$s     & 5760                            & 72 $\mu$s                           \\ \hline
(ns,~$10^{-4}$) Majorana qubit     & Hastings-Haah & 15 & 1012   & 4 $\mu$s     & 21840                           & 52 $\mu$s                           \\ \hline
(ns,~$10^{-6}$) Majorana qubit     & Hastings-Haah & 7  & {244}    & 2 $\mu$s     & 16416                           & 23 $\mu$s                           \\ \hline
\end{tabular}
\caption{
 	Physical resources required to implement a single logical qubit and to produce a
 	single T state with a T factory for factoring across the qubit parameter
 	examples specified in \tab{qubit-type}. 
 	For the logical qubit we also show which QEC code was selected and the corresponding code distance.
    }
	\label{tab:hardware-examples}
\end{table}

\subsection{Combining the requirements into application resource estimates}
\label{sec:putting-pieces-together}

We now outline how the tool combines the aspects described in \sec{algorithms-and-software} and \sec{hardware-and-QEC} to estimate thew physical resources required our three application examples with our six qubit parameter examples.
According to \sec{algorithms-and-software}, the tool provides detailed estimates of the ISA-level executable, including the number of logical qubits, the minimum number of logical time steps, and also number of logical T states required. 
The tool also estimates the maximum logical error rates for logical qubits and logical T states that achieve the required algorithm execution accuracy. 

Then, according to \sec{hardware-and-QEC}, the tool selects the QEC code and code distance for logical qubits and the T factory configuration which have the smallest physical space-time footprints while achieving the required logical error rate and logical T state error rate respectively.
This choice depends on the physical qubit parameters, and sets the number of physical qubits per logical qubit and per T factory, and also the duration of a logical time step and the T factory.
Next, the tool selects the number of T factories to balance the rate of production of T states against the rate of T state consumption demanded by the algorithm\footnote{
In some cases, the number of T factories required to meet demand may be very large, prompting the exploration of a trade-off by slowing down the algorithm allowing for a reduced number of T factories and lower qubit count.
}. 
Finally, the tool computes the total number of physical qubits as the sum of physical qubits required to implement all logical qubits and all T factories. 
It computes the overall computation time using the number of logical time steps, and the duration of a single logical time step.

\begin{figure}[t]
	\centering
	\includegraphics[width=1.0\textwidth]{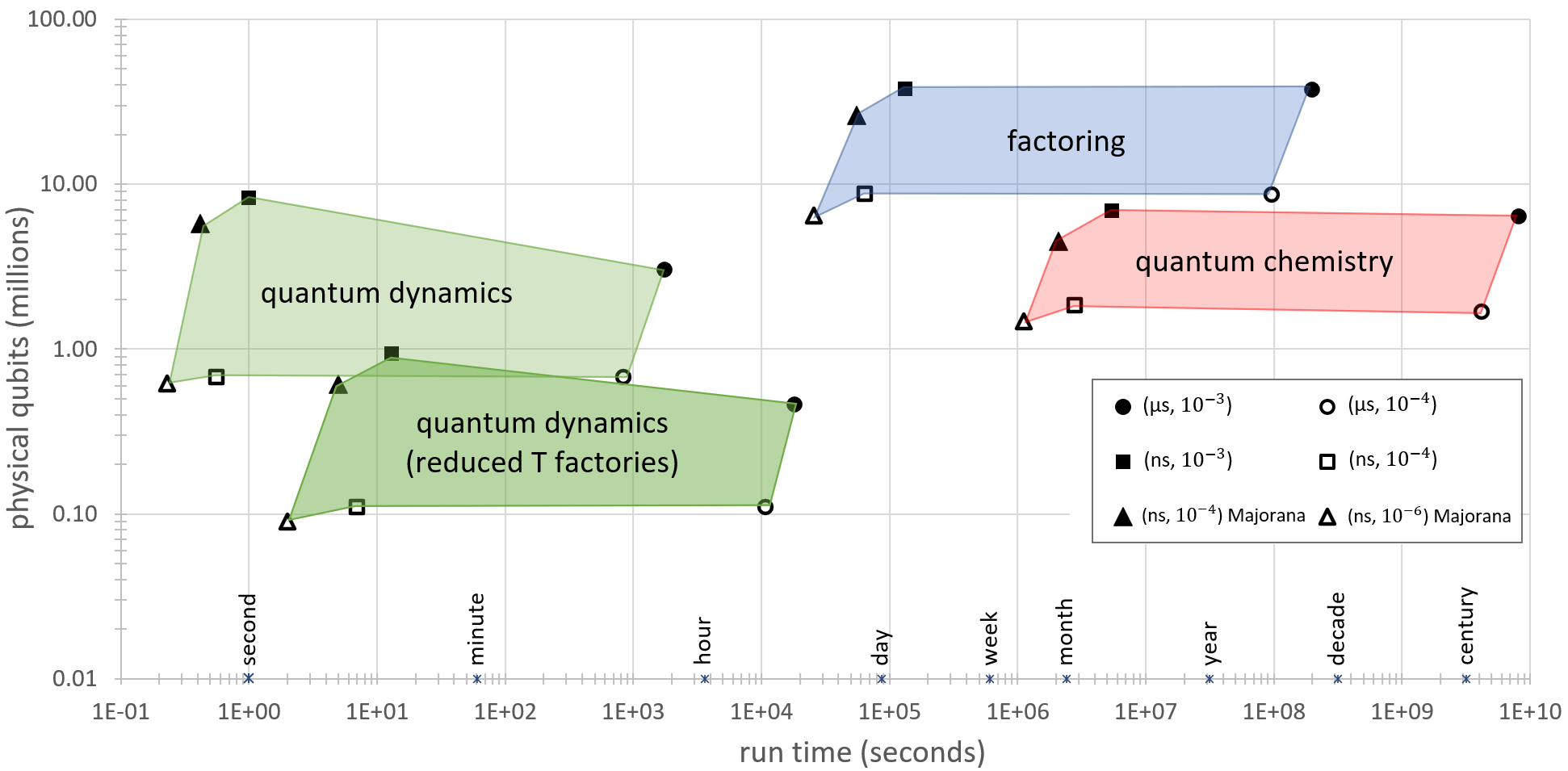}
	\caption{
	Estimates of the resources required to implement three applications, assuming the qubit parameter examples specified in \tab{qubit-type}.
	We explore a trade-off in the quantum dynamics application by considering two implementations: one which uses sufficient T factories to supply the needs of the shortest-depth algorithm and another which slows the algorithm down, allowing for a reduced number of T factories.
	}
	\label{fig:resources-chart}
\end{figure}

As can be seen in \fig{resources-chart}, qubit counts and runtimes vary by orders of magnitude based on the physical qubit parameters for all three applications.
Even the smallest practical applications such as quantum dynamics involve devices with more than 100K physical qubits and applications in chemistry and factoring require upwards of 1M qubits.
Unsurprisingly, the three orders of magnitude longer operation time for the ($\mu$s,~$10^{-3}$) and ($\mu$s,~$10^{-4}$) qubit models compared with the other models results in about three orders of magnitude longer application run times.
For quantum chemistry, this leads to impractical runtimes of more than a century. 

To build the tool and obtain these estimates we have made many choices including what algorithmic, compilation and QEC options to include and also what approximations and assumptions to make.
Other resource estimation works have made different choices, which leads to small differences in estimates~\cite{Gidney2021}.
In the appendices, we point out more explicitly these choices, assumptions and approximations, and in particular, collect together our primary assumptions in \app{primary-assumptions}.
We anticipate that future work will extend and improve the tool and framework in two ways. 
Firstly, the estimates for a given stack will become more accurate as assumptions and approximations are honed and made more realistic, for example by including more detailed and nuanced noise models.
Secondly, the estimates will become more favorable as improved solutions and optimizations are included in the stack and resource models, such as algorithmic improvements and hand-optimized compilation of important subroutines. 
We expect the broad conclusions that we draw from these results in the next section to hold true despite these choices, approximations and assumptions.
This is because our conclusions are relatively insensitive to order of magnitude changes in resource estimates.

\section{Technological implications and conclusions}
\label{sec:conclusions}

 More than two decades ago, DiVincenzo~\cite{Divincenzo2000} specified a set of fundamental requirements that any usable quantum computer should satisfy. 
 For example, DiVincenzo identified the necessity of low error rates, by requiring \textit{long  relevant  decoherence times,  much  longer  than the gate operation time}. 
 Since then, a variety of qubit technologies that satisfy DiVincenzo's criteria have been developed, including technologies such as superconducting and trapped ion qubits. 
 However, it is an open question as to what additional conditions beyond DiVincenzo's criteria a qubit technology must satisfy to scale to practical quantum advantage, and whether or not current technologies are on the path to do so.
 
 Our work sheds light on this question, by estimating and analyzing the resources needed for quantum applications with practical value implemented on fault-tolerant architecture designs assuming a variety of underlying qubit technologies. Our analysis rests on several assumptions including qubit technology trends, QEC schemes, instruction sets, and choices across other parts of the stack.  
 We have attempted to reflect the community's current best understanding of the quantum stack and its future evolution in these assumptions. Our resource estimates indicate that while scaling requirements to achieve the first scientifically interesting demonstrations of quantum advantage are not too stringent,  scaling to commercially valuable applications, such as those using computational catalysis in chemistry, presents significant challenges. 
 We deduce based on the results that to overcome these challenges and achieve scale, a qubit technology will need to be:
 
{\bf Controllable.---}
While most quantum hardware is controllable at small qubit counts, scaling qubit control to millions of fast and reliable qubit operations without introducing new noise channels is important.
Controllability is at the core of scaling up, and has implications on several aspects of the physical qubit device, including its connectivity, error rates, operation speed and size.

Regarding connectivity, it has been shown that 2D connectivity is sufficient to implement fault-tolerant quantum error correction and build logical qubits~\cite{Fowler2012,Paetznick2022,svore2005}, 
however such an array may present challenges for control and readout wiring. 
In contrast, while 1D connectivity is simpler for control wiring, it requires prohibitively low error rates \cite{jones2018} and seems unlikely to be viable with fabrication defects which permanently eliminates some qubits and operations. 
All-to-all connectivity is challenging to realize in many architectures, and 
may not be practically beneficial for large-scale implementations unless QEC schemes with good fault tolerant logical operations are developed to take advantage of such properties.

At the heart of fault-tolerant QEC is the requirement that the error rate of each physical operation (including qubit preparations, measurements, and gates) is below a threshold value.
With no connectivity restrictions, the highest known threshold~\cite{knill2005} is 3\%, and 1\% can be achieved with 2D connectivity~\cite{Fowler2012,Paetznick2022}.
The qubit overhead of QEC goes down significantly as error rates are reduced further below this threshold value; however the overheads may still be prohibitive if error rates are not an order of magnitude better, say below around 0.1\%. 
Even so, error rates have to be well below threshold while allowing for \textit{parallel} operations at a scale of \textit{millions of qubits}. 



Controllability means being able both to  control those millions of parallel operations with the desired error rates, and to readout out those millions of qubits in parallel to enable decoding of the errors at speed; all while ensuring the overarching logical clock time is fast enough to complete the computation within a month runtime or less.
To execute syndrome measurements on these qubits and communicate the quantum measurements to the decoder, we require large quantum-classical bandwidth and processing power for decoding. The exact estimates of bandwidth requirements depend on the choice of QEC code, system size and physical operation times, but roughly, for system with a few million qubits, we estimate that several terabytes per second of bandwidth will be required beteween the quantum and classical plane. Furthermore, processing these measurements at a rate that is sufficient to effectively correct errors demands petascale classical computing resources that are tightly integrated with the quantum machine.

{\bf Fast.---}
The run times identified in our study help determine the desired speed of physical operations for a given quantum architecture to be practical. 
We find that logical gate times under 10 $\mu$s, in turn requiring physical gate times around 100 ns, would be needed to complete the quantum chemistry algorithm within a month, using a few million physical qubits. 
While these gate times are realistic for solid state or photonic qubits \cite{bombin2021a}, this is several orders of magnitude faster than current 
proposals for large ion trap or neutral atom-based devices \cite{blueprintTrappedIon,Saffman_2016}.
For some algorithms, it may be possible to compensate for slower qubit speeds by using parallelization techniques, most likely resulting in increasing the required number of physical qubits.

{\bf Small.---}
Bringing together a million and more physical qubits along with their control and readout systems presents another challenge: the size of the quantum computer. 
We foresee a monolithic, single-wafer approach as desirable and estimate that a linear extent of a micron or more is needed for wiring, defining a `sweet spot' for the size of a qubit of around 10 microns~\cite{Franke2019}.

To understand this sweet spot, consider estimating the approximate footprint of the quantum plane by multiplying the number of physical qubits by the size of a physical qubit, including control electronics. 
While precise estimates on the size of leading qubit platforms are not available in the literature, we estimate that a system with 10M superconducting transmon qubits would require an area of $\sim 10$ square meters based on current designs~\cite{arute2019}. 
At such scale, a dilution refrigerator that provides the operating environment for the qubit plane will no longer be able to accommodate a monolithic quantum module. 
While modular quantum architectures that span multiple dilution refrigerators have been proposed \cite{Bravyi2022}, the reliability of the components, such as high bandwidth interconnects and low latency coherent quantum networks with associated distributed fault-tolerance schemes, and the ability to scale up while maintaining low error rates is not yet known. 
In contrast, the pursuit of a monolithic architecture in which ten million physical qubits must fit onto a single wafer suggests that shrinking the qubit to a linear extent of less than 100 microns would be needed.
This presents another set of challenges, including not making the qubit too small. If qubits are too tightly packed, then scalable control and readout of arrays of qubits will become hard. 

While no qubit technology currently implemented satisfies all of these requirements, two architecture proposals, while not yet implemented, appear to satisfy the requirements from a design perspective.  
These include recent proposals of electro-acoustic qubits \cite{Guillaud2019,Chamberland2022a}, and the topological qubit approach based on Majorana Zero Modes \cite{Karzig2017}.
Understanding these requirements on the path to scaling up, for these and other existing technologies, illuminates the opportunities and advancements needed to achieve practical quantum advantage.
It will be through studying the tradeoffs across the stack, from the algorithm to the qubit device, and then verifying them in implemnetation, that we will accelerate the development of scalable quantum computers. 
Both the framework and Azure Resource Estimator tool pave the way for the community to better understand the path to scaling up, the dependencies across the stack on the ultimate qubit numbers and runtimes, and the opportunities to bring down the costs at every layer. 
Practical quantum advantage is on the horizon; it may be accelerated through breakthrough techniques, including improved quantum error correction protocols, faster and more accurate decoders enabling improvements in the thresholds, richer logical operations and logical connectivity, improved algorithm compilation and optimization, and the discovery of low-cost, useful quantum algorithms with the promise of advantage.  These research directions, and more, can be studied in the context of resource estimation, and just may be key to unlocking quantum at scale.

\section*{Acknowledgements}
We thank Bela Bauer, Nicolas Delfosse, Alan Geller, Jeongwan Haah, Thomas H\"aner, Matthew Hastings, Christina Knapp, Chetan Nayak, Adam Paetznick and Marcus da Silva for discussions.

\newpage
\section*{Appendices}
\appendix
In these appendices, we provide a self-contained, more detailed description of how the resource estimates presented in \sec{examples} of the main text are obtained.
Since these estimates are calculated using the first release version of the Azure Quantum Resource Estimator, hereafter called `the tool', these appendices also serve to specify the choices, assumptions and approximations of the tool.

\begin{figure}[h]
	\centering
	\includegraphics[width=0.9\textwidth]{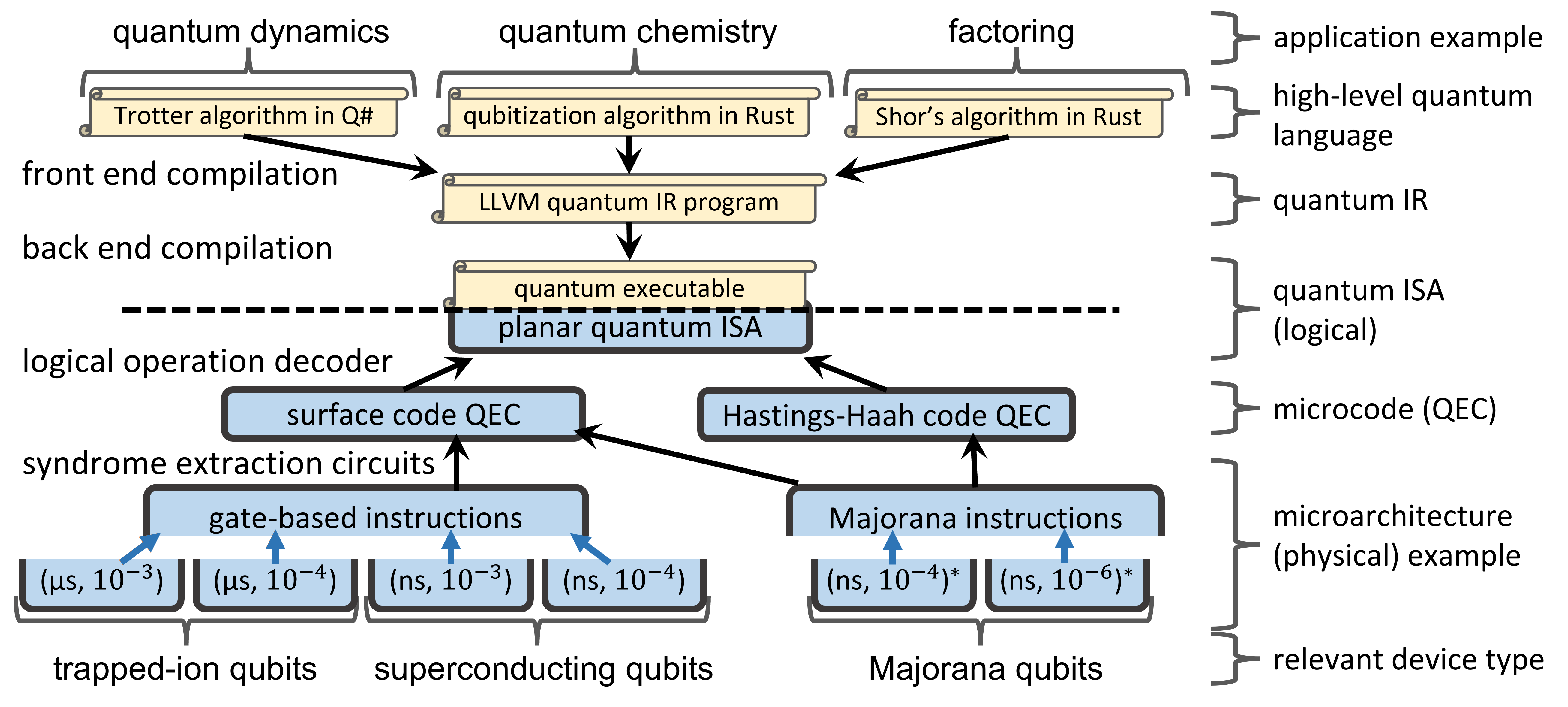}
	\caption{
The examples and options included in the quantum stack for our resource estimates.
We consider three application examples at the top, and a range of hardware parameter examples relevant for a variety of hardware approaches at the base of the stack. 
We label layers of the quantum stack on the right, and the maps between these layers on the left.
To highlight our qubit parameter examples, we separate the microarchecture layer into two sub-layers.
All our examples flow through the same planar quantum ISA.  
Applications are translated down the software stack and the resulting ISA-level executable is input to the tool. 
The tool also takes as input configurable architecture models that include fault-tolerance details and physical qubit models. 
The tool outputs resource estimates such as the number of physical qubits or time required to run the application.
	}
	\label{fig:examples}
\end{figure}

\fig{examples} summarizes the set of examples and major options that form the stacks that we model to estimate resources, and \tab{physical-and-logical-overhead} collects together the input, output, and intermediate data used to calculate the estimates for each application example using each qubit parameter example, which are used to produce \fig{resources-chart} in the main text.
The appendices methodically address each aspect of this estimation calculation.

While it is impossible to fully model every aspect of a future large-scale quantum computing system, here we strive to model system components that are expected significantly impact performance; we abstract away or neglect aspects which are expected to have small contributions to the cost.
In this tool we have made a number of assumptions to enable simple and efficient, yet effective, resource estimation.  
The formulas used in our estimates are expressed to the lowest order in parameters which are expected to be small (such as error rates).
These formulas are therefore only valid for sufficiently small values of those parameters.
(Various checks are included in the tool to flag when this is not the case given user inputs.)
We anticipate that the resource estimates we present here will be further improved by incorporating additional details and relaxing a number of our assumptions. 
The primary assumptions that are made throughout the various aspects of this analysis are collected together and listed in \app{primary-assumptions}.

The remaining appendices are divided into three parts. 
The first part, \app{noise}-\app{cat-state}, describes models for different components of a large-scale quantum computer. 
In \app{resource-estimates}, we use these models to estimate physical resources (qubit counts, runtime) required for executing quantum applications. 
The third part, \app{applications}, details the applications and algorithmic optimizations used to reduce resource requirements. 

\begin{table}[h]
	\begin{tabular}{|c||c|c|c|c||c||c|c|c||c|c|}
		\hline
		  \multirow{4}{*}{\textbf{application}} &   \multicolumn{4}{|c||}{planar quantum ISA} &  &  \multicolumn{3}{|c||}{QEC} & \multicolumn{2}{|c|}{\textbf{resource}}   \\
		 &   \multicolumn{4}{|c||}{requirements} & \textbf{qubit} &  \multicolumn{3}{|c||}{optimization} & \multicolumn{2}{|c|}{\textbf{requirements}}  \\
		\cline{2-5}\cline{7-11}
		\multirow{2}{*}{} &  \multirow{2}{*}{$\Q$} & \multirow{2}{*}{$\TimeLogicalFast$} & { \color{gray} \multirow{2}{*}{$\TimeLogical$}} & \multirow{2}{*}{$\R$} & \textbf{parameters}  & \multirow{2}{*}{$d$} & \multirow{2}{*}{$\nFactories$} & factory  & \textbf{physical}  & \textbf{physical}   \\
		 &  & &  &    &     & && ratio & \textbf{qubits}  & \textbf{run time}  \\
		\hline
		\hline
		&   &  &  {\color{gray} $1.5 \cdot 10^{5}$} & & \boldmath{$(\mu \textbf{s},~10^{-3})$ } & 19 & 199 & 95\%  &  \textbf{3.0M}  & \textbf{29 mins} \\
		&   &  & {\color{gray} $1.5 \cdot 10^{5}$} & & \boldmath{$(\mu \textbf{s},~10^{-4})$ } & 9 & 199 & 95\%  &  \textbf{0.68M}  & \textbf{14 mins} \\
		\textbf{quantum} &   \multirow{2}{*}{230} & \multirow{2}{*}{$1.5 \cdot 10^{5}$} & {\color{gray} $1.5 \cdot 10^{5}$} & \multirow{2}{*}{$2.4 \cdot 10^{6}$}    & \boldmath{$(\textbf{ns},~10^{-3})$ } & 19 & 242 & 98\%  &  \textbf{8.2M}  & \textbf{1.1 secs}  \\
		\textbf{dynamics} && & {\color{gray} $1.5 \cdot 10^{5}$} &    & \boldmath{$(\textbf{ns},~10^{-4})$ } & 9 & 199 & 95\%  &  \textbf{0.68M}  & \textbf{0.56 secs}   \\
		 &   & & {\color{gray} $1.5 \cdot 10^{5}$} &  &  \boldmath{$(\textbf{ns},~10^{-4})^*$} & 9 & 260 & 99\%  &  \textbf{5.8M}  & \textbf{0.42 secs} \\
		 &   & & {\color{gray} $1.5 \cdot 10^{5}$} &  &  \boldmath{$(\textbf{ns},~10^{-6})^*$} & 5 & 224 & 95\%  &  \textbf{0.62M}  & \textbf{0.23 secs} \\
		\hline
		\hline
		&   &  &  {\color{gray} $1.5 \cdot 10^{6}$} & & \boldmath{$(\mu \textbf{s},~10^{-3})$ } & 21 & 18& 56\%  &  \textbf{0.46M}  & \textbf{5.3 hours} \\
		\textbf{quantum} &   &  & {\color{gray} $1.5 \cdot 10^{6}$} & & \boldmath{$(\mu \textbf{s},~10^{-4})$ } & 11 & 17& 50\%  &  \textbf{0.11M}  & \textbf{2.8 hours} \\
		\textbf{dynamics} &   \multirow{2}{*}{230} & \multirow{2}{*}{$1.5 \cdot 10^{5}$} & {\color{gray} $1.5 \cdot 10^{6}$} & \multirow{2}{*}{$2.4 \cdot 10^{6}$}    & \boldmath{$(\textbf{ns},~10^{-3})$ } & 21 & 22 & 78\%  &  \textbf{0.94M}  & \textbf{13 secs}  \\
		\textbf{(reduced} && & {\color{gray} $1.5 \cdot 10^{6}$} &    & \boldmath{$(\textbf{ns},~10^{-4})$ } & 11 & 17& 50\%  &  \textbf{0.11M}  & \textbf{6.7 secs}   \\
		 \textbf{T factories)} &   & & {\color{gray} $1.5 \cdot 10^{6}$} &  &  \boldmath{$(\textbf{ns},~10^{-4})^*$} & 11 & 22 & 79\%  &  \textbf{0.61M}  & \textbf{5.0 secs} \\
		 &   & & {\color{gray} $1.5 \cdot 10^{6}$} &  &  \boldmath{$(\textbf{ns},~10^{-6})^*$} & 5 & 23 & 66\%  &  \textbf{0.09M}  & \textbf{2.3 secs} \\
		\hline
		\hline
		&  & & {\color{gray} $4.1 \cdot 10^{11}$} &  &  \boldmath{$(\mu \textbf{s},~10^{-3})$ } & 33 & 15 & 6.9\%  &  \textbf{6.4M}  & \textbf{260 years} \\
		& &  & {\color{gray} $4.1 \cdot 10^{11}$}  &  &  \boldmath{$(\mu \textbf{s},~10^{-4})$ } & 17 & 14 & 5.9\%  &  \textbf{1.6M}  & \textbf{130 years} \\
		\textbf{quantum} &   \multirow{2}{*}{2740} & \multirow{2}{*}{$4.1 \cdot 10^{11}$} & {\color{gray} $4.1 \cdot 10^{11}$} & \multirow{2}{*}{$5.4 \cdot 10^{11}$}    & \boldmath{$(\textbf{ns},~10^{-3})$ } & 33 & 17 & 14\%  &  \textbf{6.9M}  & \textbf{2.0 months}  \\
		\textbf{chemistry} && & {\color{gray} $4.1 \cdot 10^{11}$} &    & \boldmath{$(\textbf{ns},~10^{-4})$ } & 17 & 17 & 15\%  &  \textbf{1.9M}  & \textbf{1.0 month}   \\
		 &   & & {\color{gray} $4.1 \cdot 10^{11}$} &  &  \boldmath{$(\textbf{ns},~10^{-4})^*$} & 17 & 19 & 22\%  &  \textbf{4.5M}  & \textbf{24 mins} \\
		 &   & & {\color{gray} $4.1 \cdot 10^{11}$} &  &  \boldmath{$(\textbf{ns},~10^{-6})^*$} & 9 & 19 & 22\%  &  \textbf{1.3M}  & \textbf{12 days} \\
		\hline
		\hline
		&   &  & {\color{gray} $1.2 \cdot 10^{10}$} &  &  \boldmath{$(\mu \textbf{s},~10^{-3})$ } & 27 & 13 & 0.6\%  &  \textbf{37M}  & \textbf{6.2 years} \\
		&   & & {\color{gray} $1.2 \cdot 10^{10}$} &  &  \boldmath{$(\mu \textbf{s},~10^{-4})$ } & 13 & 14 & 0.8\%  &  \textbf{8.6M}  & \textbf{3.0 years} \\
		\multirow{2}{*}{\textbf{factoring}} &   \multirow{2}{*}{25481} & \multirow{2}{*}{$1.2 \cdot 10^{10}$} & {\color{gray} $1.2 \cdot 10^{10}$} & \multirow{2}{*}{$1.5 \cdot 10^{10}$}    & \boldmath{$(\textbf{ns},~10^{-3})$ } & 27 & 15 & 1.3\%  &  \textbf{37M}  & \textbf{1.5 days}  \\
		 && & {\color{gray} $1.2 \cdot 10^{10}$} &    & \boldmath{$(\textbf{ns},~10^{-4})$ } & 13 & 18 & 1.1\%  &  \textbf{8.7M}  & \textbf{18 hours}   \\
		 &   & & {\color{gray} $1.2 \cdot 10^{10}$} &  &  \boldmath{$(\textbf{ns},~10^{-4})^*$} & 15 & 15 & 0.9\%  &  \textbf{26M}  & \textbf{15 hours} \\
		 &   & & {\color{gray} $1.2 \cdot 10^{10}$} &  &  \boldmath{$(\textbf{ns},~10^{-6})^*$} & 7 & 13 & 1.2\%  &  \textbf{6.2M}  & \textbf{7.1 hours} \\
		\hline
	\end{tabular}
	\caption{
    Estimated resource costs to implement some example applications using hardware devices with various parameters.
	The application is compiled down from a high-level quantum program to a program expressed in the planar quantum ISA, with parameters as shown.
	To supply the planar quantum ISA with the capability to run this program, QEC is used. 
	The code distance $d$ of the error correcting code and the number of T state distillation factories $\nFactories$ are carefully optimized to minimize the resources needed while providing enough fault protection that the overall computation runs reliably. We also show the percentage of the physical qubits which are used for T state distillation factories.
	We use an asterisk to indicate Majorana qubits.
	In the second implementation of the quantum dynamics example we slow the algorithm down by setting $\TimeLogical = 10 \TimeLogicalFast$, which allows for a reduction in the number of T factories, thereby reducing the qubit overhead.
	}
	\label{tab:physical-and-logical-overhead}
\end{table} 

\section{Physical qubit models}
\label{app:noise}

We begin by specifying the instruction sets and noise models for the physical qubits representation in our analysis, at the level of the microarchitecture in our quantum stack.
We also provide a set of example qubit parameters.

\begin{figure}[h]
	\centering
	\includegraphics[width=0.8\textwidth]{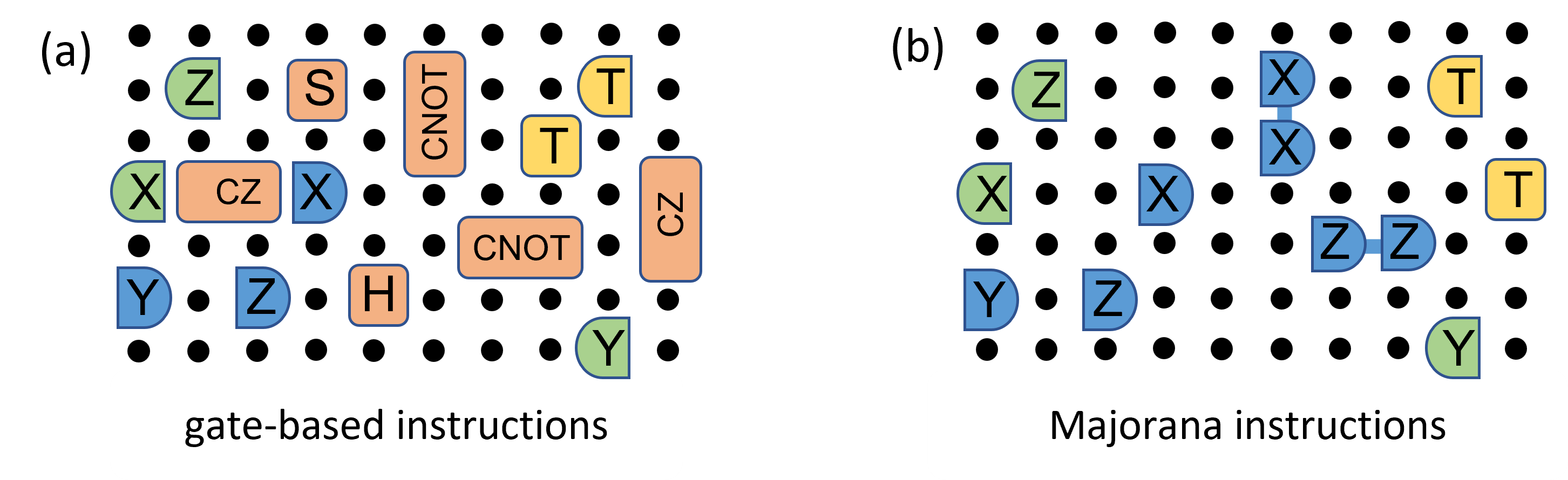}
	\caption{%
	We consider gate-based and Majorana instruction sets for physical qubits. 
	Both sets include state preparation in the Pauli basis (green), 
	T state preparation (yellow), T gate application (yellow), and measurement in all Pauli basis (blue). 
	(a) The gate-based instruction set also includes Hadamard H and S gates, and CNOT and CZ entangling gates between adjacent qubits (all orange).
	(b) The Majorana instruction set also allows non-destructive joint Pauli measurements of adjacent qubits (blue).
	}\label{fig:physical-operations}
\end{figure}

We assume that physical qubits are capable of implementing one of two instruction sets; see \fig{physical-operations}.
The \emph{gate-based} instruction set implements unitary operations such as CNOT or CZ as its native entangling operations. 
On the other hand, the \emph{Majorana} instruction set implements parity checks such as ZZ and XX measurements as its native entangling operations.\footnote{Note that in some literature, the Majorana instruction set is known as `measurement-based', but this can cause confusion with quantum computing using cluster states, which is also commonly referred to as measurement-based.} 
With Majorana instruction sets, unitary operations are realized using a sequence of measurements.
For both gate-based and Majorana instruction sets in \fig{physical-operations}, we specify the allowed operations by visualizing the qubits laid out in a array with nearest neighbor connectivity.
This visualization can either represent the true layout of the qubits in the hardware, such as for solid-state based approaches or can be considered as no more than a visualization to ensure that all operations in the instruction set are available, such as for trapped ions. 
Note that a particular hardware approach may have other allowed operations not in one of these instruction sets --- these additional operations are not relevant for our analysis as we assume only operations from the instruction sets are used.  

We assume the standard noise model known as \emph{circuit noise}, wherein each operation on the physical qubits fails independently.
More specifically, we assume each Clifford operation fails with probability $p$, where Clifford operations include the Hadamard and phase gates, the Controlled NOT (CNOT) and Controlled Z (CZ)  gates, single-qubit basis state preparation, and one-qubit and two-qubit Pauli measurements.
We consider an idle qubit for a single time step to be an identity operation which can also fail with probability $p$.
The preparation of a single-qubit T state, which is a non-Clifford operation, is assumed to fail with probability $p_T$.
Failures are modeled by introducing random Pauli operators on qubits targeted by the operation.
When the operation which fails is a measurement, the outcome is randomly flipped in addition to applying a Pauli operator to the support of the measurement.
We also use the terminology of error rate in place of infidelity when describing noisy states, by making the simplifying assumption that the noise on the state arose from circuit noise (or logical Pauli noise where appropriate) with a given error rate.

In \tab{qubit-type}, we consider a range of example qubit parameters which abstractly represent different regimes of operation fidelity and time. 
These values of parameters are selected to be relevant for different hardware approaches. 
For readability, we label each example by the approximate operation time, Clifford operation error rate and instruction set. 
For example, the first qubit has a gate time $t_\text{gate}$ and measurement time $t_\text{meas}$ of 100 $\mu$s, a Clifford error rate $p$ of $10^{-3}$ with a gate-based instruction set; therefore, we label it ($\mu$s, $10^{-3})$. 
The first four examples are gate-based qubits and the remaining two are Majorana qubits. 
To fit the simple circuit noise model which we assume, we choose a uniform error rate for all Clifford operations, which can be assumed to equal the highest of the error rates for any Clifford operation in the instruction set (including single-qubit gates and measurements and two-qubit gates and measurements). 
We select the parameters for each example as follows:
\begin{itemize}
    \item \textbf{$(\boldsymbol{\mu}$s,~$\mathbf{10^{-3}}$) qubit} and \textbf{$(\boldsymbol{\mu}$s,~$\mathbf{10^{-4}}$) qubit} may be relevant for future versions of trapped ion qubits~\cite{Cirac1995,pino2021}, which typically have operations times in the microsecond regime. 
    We choose 100 $\mu$s as the gate and measurement time based on typical assumptions for ion qubits~\cite{Leung2018, Murali2020}. 
    We assume high-quality single qubit gates with an error rate of $10^{-6}$ and two cases for two-qubit gates with error rates of $10^{-3}$ and $10^{-4}$ based on the architectural assumptions in Ref.~\cite{blueprintTrappedIon}.

    \item \textbf{(ns,~$\mathbf{10^{-3}}$) qubit} and \textbf{(ns,~$\mathbf{10^{-4}}$) qubit} may be relevant for future versions of superconducting transmon qubits~\cite{transmonSchoelkopf,schreier2008,kjaergaard2020,arute2019,hong2020,Steffen2011} or spin qubits~\cite{kane1998,Hanson2007,Jnane2022}, which typically have operation times in the nanosecond regime. 
    We project 50 ns gate times and 100 ns measurement times for future systems based on recent system implementations~\cite{googleSurfaceCode, ibmMontreal}, which can be viewed as relatively optimistic (particularly for measurements).  
    We evaluate two cases, with $10^{-3}$ and $10^{-4}$ two-qubit gate error rates, respectively, as realistic and optimistic targets for a scaled up system~\cite{ibmMontreal, Gidney2021}.

    \item \textbf{(ns,~$\mathbf{10^{-4}}$) Majorana qubit} and \textbf{(ns,~$\mathbf{10^{-6}}$) Majorana qubit} may be relevant for future Majorana qubits~\cite{kitaev2001,sarma2015,Karzig2017}. 
    For these qubits, we assume that measurements and the physical T gate each take $100$ ns. 
    Owing to topological protection in the hardware, we assume single and two-qubit measurement error rates (Clifford error rates) of $10^{-4}$ as a realistic target and $10^{-6}$ as an optimistic target. Non-Clifford operations in this architecture do not have topological protection, so we assume a $5\%$ and $1\%$ error rate for non-Clifford physical T gates for the realistic and optimistic models respectively.
\end{itemize}

Together, these examples cover a range of operation times and error rates, enabling sufficient exploration of the resource costs  anticipated to enable practical quantum applications.
Note that in \fig{examples} and \tab{physical-and-logical-overhead}, we use an asterisk to indicate the Majorana qubit for notational convenience.

\section{Quantum error correction and the planar quantum ISA}
\label{app:logical-operations-details}

Here we review two quantum error correction (QEC) schemes, focusing on estimates of the resources required for their implementation and for applying fault-tolerant logical operations on the encoded information.
For qubits with a gate-based instruction set, we assume the surface code. 
It is the best-understood QEC scheme for this class of qubits and offers a high threshold for practical implementation. 
For qubits with a Majorana instruction set, we consider both the surface code and also the Hastings-Haah code, which is a recently developed QEC scheme that offers better space-time costs than surface codes on Majorana qubits in many regimes~\cite{Hastings2021,Paetznick2022}.
It is also in principle possible to implement the Hastings-Haah code with qubits that use a gate-based instruction set, but the overhead is higher than the surface code in all regimes of interest in this case so we do not include it~\cite{Gidney2022}.

In \tab{error-correction} we provide formulas to estimate the physical qubit and time overheads required to implement logical qubits with the surface code and the Hastings-Haah code as a function of qubit design parameters. 
We estimate the logical error rate $\Plog(d)$ of a patch of surface code or Hastings-Haah code of distance $d$ with physical error rate $p$ using the formula
\begin{eqnarray}
\label{eq:logical-error-rate-formula}
\Plog(d) = a \left( \frac{p}{p^*}\right)^{\frac{d+1}{2}},
\end{eqnarray}
where the pre-factor $a$ and threshold value $p^*$ can be extracted numerically from simulations.

\begin{table}[h]
	\begin{tabular}{|c|c|c|c|}
		\hline
		QEC scheme		&	logical error rate		& qubits per logical qubit &  logical time step \\
		\hline
		\hline
		surface code	& \multirow{2}{*}{$\Plog_{\text{sur}}(d) = 0.03 \left(\frac{p}{0.01}\right)^{\frac{d+1}{2}}$} &	\multirow{2}{*}{$n_{\text{sur}}(d) = 2d^2$}		& \multirow{2}{*}{$\tau_{\text{sur}}(d) =(4 \tGate + 2 \tMeas) d$} \\
		(gate-based qubits) &&& \\
		\hline
		surface code	& \multirow{2}{*}{$\Plog_{\text{sur,meas}}(d) = 0.08 \left(\frac{p}{0.0015}\right)^{\frac{d+1}{2}}$} &	\multirow{2}{*}{$n_{\text{sur,meas}}(d) = 2d^2$}		& \multirow{2}{*}{$\tau_{\text{sur,meas}}(d) = 20 \tMeas d$} \\
		(meas-based qubits) &&& \\
		\hline
		Hastings-Haah code	& \multirow{2}{*}{$\Plog_{\text{HH}}(d) = 0.07 \left(\frac{p}{0.01}\right)^{\frac{d+1}{2}}$}	&	\multirow{2}{*}{$n_{\text{HH}}(d) = 4d^2 + 8(d-1)$}	& \multirow{2}{*}{$\tau_{\text{HH}}(d) = 3 \tMeas d$} \\
		(meas-based qubits) &&& \\
		\hline
	\end{tabular}
	\caption{
	Options of error correction schemes.
	We model each logical qubit as having a probability $\Plog$ of failing during a single logical time step, which can be tuned by changing the code distance $d$, which is an odd integer. 
	Increasing $d$ increases the number of qubits $n(d)$ required to store each logical qubit and the time $\tau(d)$ required for each logical time step.
	The formulas are based on Refs.~\cite{Fowler2012} and \cite{Wang2011} for gate-based surface codes, on Refs.~\cite{Tran2020} and \cite{Chao2020} for surface codes implemented with Majorana instruction sets (replacing 8 steps to measure a single stabilizer in  Ref.~\cite{Tran2020} by 20 steps to measure all stabilizers), and on Ref.~\cite{Paetznick2022} for Hastings-Haah codes.
    }
	\label{tab:error-correction}
\end{table}

\begin{figure}[t]
	\centering
	\includegraphics[width=1.0\textwidth]{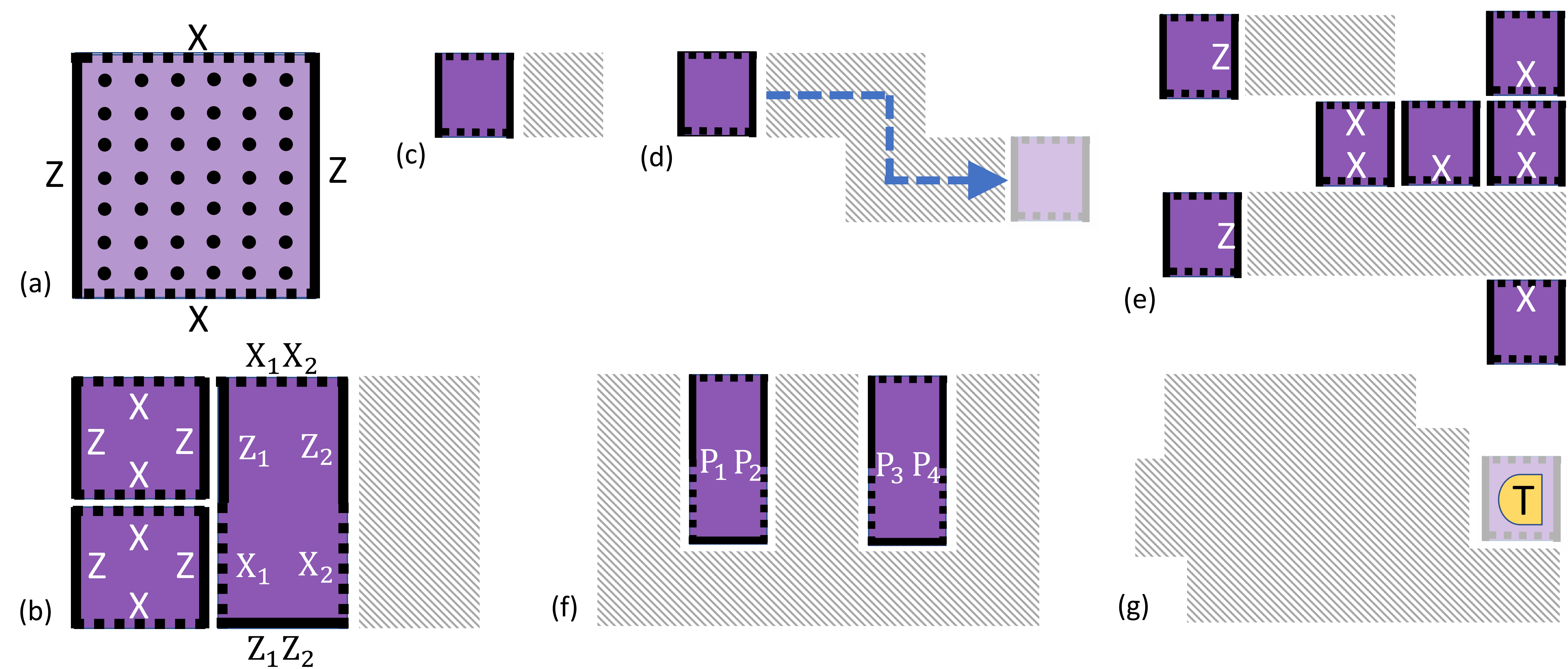}
	\caption{%
	\textbf{Fault-tolerant logical operations: instruction set of the planar quantum ISA}.
	(a) Patches of physical qubits encode logical qubits in either the surface code or the Hastings-Haah code, with error correction performed using operations from the instruction set.
	We group the array of physical qubits into tiles, each containing $n(d)$ physical qubits, and count logical operations in units of logical time steps $\tau(d)$ as defined in \tab{error-correction}.
	(b) We focus here on \textit{one-tile patches}, and \textit{two-tile patches}, which encode one and two qubits into $n(d)$ and $2n(d)$ physical qubits respectively. 
	We label sections of the boundary of these patches which are relevant for the logical operations of the planar quantum ISA. 	
	\textbf{Preparation} of qubits in one-tile and two-tile patches in computational basis states can be applied in $0\tau(d)$.
	Arbitrary \textbf{Pauli unitaries} can be applied to qubits in one-tile and two-tile patches in $0\tau(d)$.
	Destructive \textbf{measurement} of a qubit in a one-tile patch in the X or Z basis and of both of the qubits in a two-tile patch (either both in the X basis or both in the Z basis) can be done in $0\tau(d)$.
	(c) \textbf{Hadamard and phase unitaries} can be applied to a qubit in a one-tile patch with an adjacent ancilla tile in $3\tau(d)$ and $2\tau(d)$ respectively.
	The ancilla can be on any boundary for the Hadamard, and next to a Z boundary for the phase.
	(d) A \textbf{move} of a qubit in a one-tile patch to a different tile along a path connecting any boundaries of the starting and ending tiles in $1\tau(d)$.
	(e) A non-destructive \textbf{multi-qubit X, Z type measurement} of the qubits in a set of one-tile patches can be implemented in $1\tau(d)$.
	This can be done by forming a connected region of tiles, where each tile hosting a qubit to be measured in the X (Z) basis connects through one or both of its X (Z) boundaries, and connections can be between adjacent qubit tiles, or mediated by ancilla tiles. 
	(h) A non-destructive \textbf{multi-qubit arbitrary Pauli measurement} of a set of qubits stored in two-tile patches can be performed in $1\tau(d)$.
	This can be achieved with a connected region of ancilla tiles with access to all of the X and Z boundaries of each of the two-tile patches. 
	(i) \textbf{T state preparation} in either a one-tile patch or a two-tile patch can be achieved using a distillation factory as described in \app{distillation-overhead}, using a non-integer number of tiles and logical time steps which depends on the quality requirements of the state.
	}\label{fig:logical-operations}
\end{figure}

Protected logical operations can be applied to logical qubits stored in the surface code or the Hastings-Haah code.
We assume precisely the same types of logical patches, set of logical operations and costs (in units of number of tiles and logical time steps) for surface codes and Hastings-Haah codes, which forms the logical instruction set which we call the \textit{planar quantum ISA} shown in \fig{logical-operations}.
The differences that we account for between these two QEC codes are all captured by the different formulas for the logical error rate $P(d)$, number $n(d)$ of physical qubits in a tile and the time $\tau(d)$ for a logical time step for a given distance $d$, as specified in \tab{error-correction}.
The logical operations can be inexpensive, almost perfect and instantaneous (Pauli unitaries, single-qubit preparations and measurements), medium cost, requiring ancilla tiles and one logical time step (multi-qubit Pauli measurements), or expensive, such as T state preparation for the execution of so-called non-Clifford gates, which requires a significant number of time steps and auxiliary qubits, which is discussed in detail in \app{distillation-overhead}.
The patch choices and logical Clifford operations and costs are largely based on Refs.~\cite{Litinski2019,Bombin2021}.

As a simplification, we do not account for the overhead contributions associated with a number of known challenges. 
The circuits expressed in the physical-level instruction set which are used to implement QEC codes and logical operations fault-tolerantly are referred to as \textit{syndrome measurement circuits}.
For surface codes, the two-tile patches may require additional technical challenges to implement the necessary syndrome measurement circuits, such as measuring higher-weight stabilizer generators~\cite{Chamberland2021,Chamberland2022b}. 
Two-tile patches for Hastings-Haah codes have not been studied in detail in the literature but may pose similar implementation challenges.
Moreover, the size and shape of a tile may need to be altered to accommodate the non-standard patches (such as the two-tile patch) and to account for ancillas between patches that can aid lattice surgery.
We also assume that a Clifford operation which takes $A$ logical time steps to implement and involves $B$ tiles fails with probability $A \cdot B \cdot \Plog(d)$.

\section{T state distillation factories}
\label{app:distillation-overhead}

The non-Clifford T state preparation in the planar quantum ISA in \fig{logical-operations} is crucial because the other operations in that quantum ISA, which are all Clifford operations, and are not sufficient for universal quantum computation~\cite{GottesmanKnill}.
To implement non-Clifford operations for practical-scale algorithms, we require low error rate T gates or T states, however these can be difficult to directly implement on logical qubits~\cite{Bravyi2013,beverland2016a}, and can also be difficult for some physical qubits~\cite{sarma2015,Karzig2017}. 
Instead, the required low error rate T states are produced using a \emph{T state distillation factory}~\cite{knill2004,bravyi2005}, which we sometimes refer to as a `T state factory', a `distillation factory' or even just a `factory' for short.
Here we consider the resources required to implement factories which produce T states in distance-$d$ surface or Hastings-Haah codes with error rate $\PlogDis$.

The operation of a T state factory typically begins by first producing imperfect T states using some means, for example, a low-quality physical T gate acting on a physical qubit. 
These T states are then typically transferred to a logical qubit and is refined using a distillation unit, which uses only Clifford operations. 
This procedure can be iterated, where the output T states of one round are fed into the next round as inputs. 
In the error rate of physical T gates $p_T$ is much larger than that of physical Clifford operations $p$, it can make sense to do some distillation in physical qubits before transferring the state to a logical qubit.

\begin{figure}[t]
	\centering
	\includegraphics[width=0.95\textwidth]{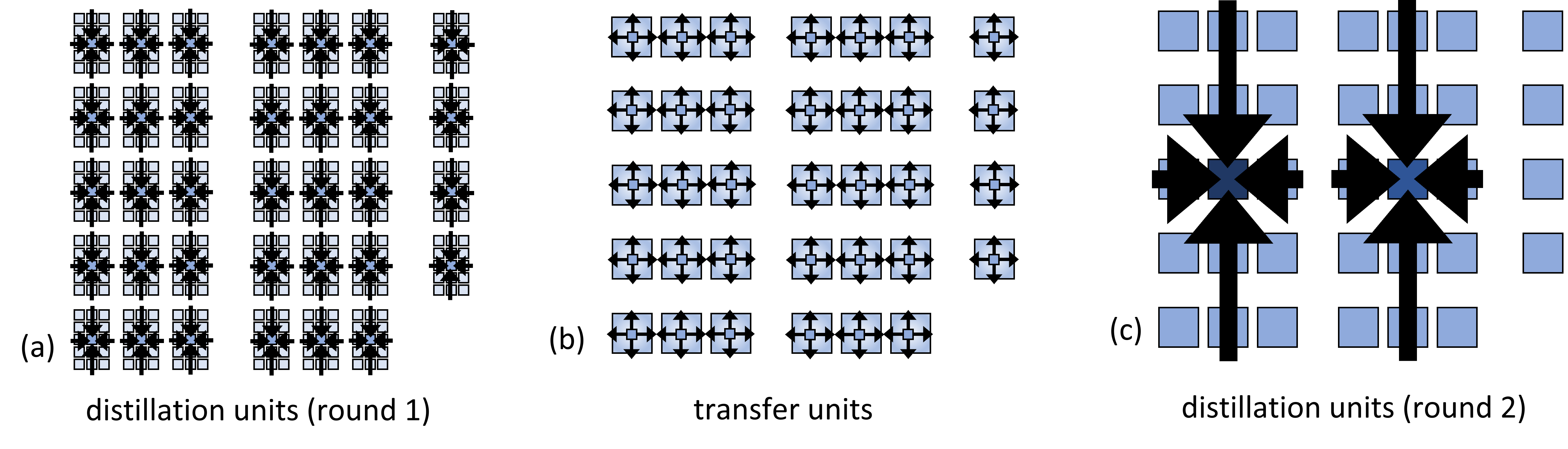}
	\caption{%
		An example of a two-round T state distillation factory. 
		First, a transfer unit encodes $c_0 = 34 \cdot 15 = 510$ noisy T states into distance-$d_1$ code patches (not shown).
		(a) Next, in the first distillation round, each of these T states are input into one of $c_1=34$ copies of a 15-to-1 distillation unit are applied on the distance-$d_1$ code patches.
		It is possible that not all of the 34 copies of the distillation unit are accepted. 
		(b) For those copies which are accepted, the code patch hosting the T state produced is expanded from distance $d_1$ to distance $d_2$, and moved into place for the next distillation round using a transfer unit.
		(c) In the second distillation round, $30$ T states (provided at least that many remain) are fed into $c_2=2$ copies of a 15-to-1 distillation unit. 
		If fewer than $30$ T states remain (but more than 14), then just one of the $c_2=2$ copies of the distillation unit can be implemented.
		The T states encoded in distance-$d_2$ codes produced by round two form the output of the distillation factory.
		}\label{fig:distillation}
\end{figure}

We consider T state distillation factories which are implemented in a sequence of rounds, where each round consists of a set of identical distillation units run in parallel; see \fig{distillation}.
We focus on factories composed of two types of units. \textit{Distillation units} (see \tab{distillation-modules}) take in a set of T states with a given quality, and output a smaller set of T states, but which typically have a higher quality.
We assume that distillation units have the same kind of qubits as input and output, for example distance-$d$ surface codes.
Distillation units typically involves a post selection test - if it is passed, then the output is accepted, otherwise the output is discarded.
\textit{Transfer units} take T states stored in qubits of one type in one location and transfer them to T states stored in qubits of a different type at another location.
For example a transfer unit could take a set of T states stored in physical Majorana qubits, and transfer them into a set of distance-$d$ Hastings-Haah codes positioned where they are needed as inputs to a distillation unit in the next round.  
Another example would take qubits in distance-$d_1$ surface codes at the locations output from a previous round of distillation, and transfer them into distance-$d_2$ surface codes at the locations where they are needed as inputs to a distillation unit in the next round.
In our analysis we neglect additional overheads required for implementing the transfer units and any noise they introduce.
We specify a distillation factory $\mathcal{D}$ by stating: the number of rounds it uses $R$, and for each round $r \in \{ 1,2,\dots R\}$, both the distillation unit $M_r$ (including the qubit type, such as a specific physical qubit example, or a logical surface code or logical Hastings-Haah code, and code distance) and the number of copies $c_r$ of that unit.
We do not specify the transfer units between rounds because we do not include the associated costs in our resource estimates.
See \tab{distillation-examples} for some example distillation factories.

\begin{table}[h]
	\begin{tabular}{|c||c|c|c|c|c|c|}
		\hline
		distillation &  \# input Ts & \# output Ts  & acceptance & \# qubits  & time & output \\
		 unit & &  & probability &  & & error-rate \\
		\hline
		\hline
		15-to-1 space-eff. & 15 & 1 & $1-15p_T - 356p$ & $12$ & $46\tMeas$ & $35 p_T^3 + 7.1p$  \\
		physical   & & & & & & \\
		\hline
	    15-to-1 space-eff. & 15 & 1 & $1-15P_T - 356P$ & $20n(d)$ & $13\tau(d)$ & $35 P_T^3 + 7.1P$  \\
	    logical  & & & & & & \\
	   	\hline
 		15-to-1 RM prep. & 15 & 1 & $1-15p_T - 356p$ & $31$ & $23\tMeas$ & $35 p_T^3 + 7.1p$  \\
 		physical   & & & & & & \\
 		\hline
 	    15-to-1 RM prep. & 15 & 1 & $1-15P_T - 356P$ & $31n(d)$ & $13\tau(d)$ & $35 P_T^3 + 7.1P$  \\
 		logical  & & & & & & \\
		\hline
	\end{tabular}
	\caption{
	Distillation units.
	Physical distillation units assumes Majorana qubits with error rates of $p$ and $p_T$ for Clifford and input T states respectively.
	Logical distillation units assume either the surface code or the Hastings-Haah code, with each logical qubit requiring $n$ physical qubits, each logical time step requiring a time $\tau$, and logical error rates of $P$ and $P_T$ for Clifford and input T states respectively.
	The units are implemented by circuits of allowed operations as shown in \fig{distillation-circuits-1} and \fig{distillation-circuits-2}.
	The coefficients for $p$ and $P$ are estimates based on numerical results for similar circuits, but may be inaccurate.
	}
	\label{tab:distillation-modules}
\end{table}

\begin{figure}[h]
	\centering
	\includegraphics[width=0.8\textwidth]{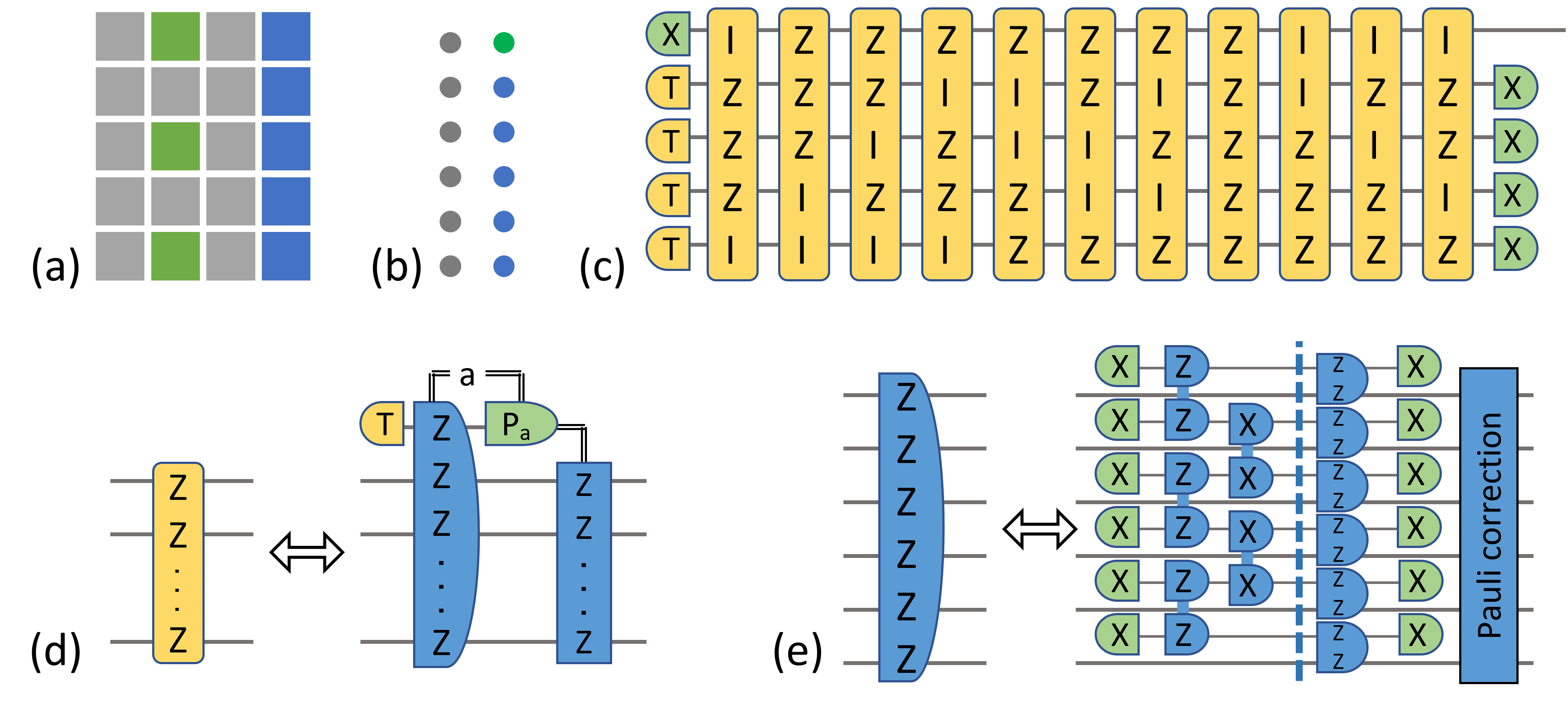}
	\caption{%
	    Space-efficient 15-to-1 distillation units.
		(a) and (b) show the logical and physical layouts using 20 patches and 12 physical qubits respectively.
		(c) A unitary version of the distillation circuit based on Ref.~\cite{Litinski2019b}. 
		Four T states are fed in at the beginning of the circuit, and then a sequence of diagonal non-Clifford rotations are applied (yellow boxes, with the rotation $\exp(-iP \frac{\pi}{8})$ applied for a box containing Pauli $P$).
		(d) A diagonal non-Clifford rotation $\exp(-iP \frac{\pi}{8})$ can be implemented using a small Clifford circuit given an ancilla prepared in the T state. 
		First, a multi-qubit Z measurement is applied with the support of $P$ extended to include the ancilla, the outcome is $a$. 
		Then the ancilla is measured in the basis $P_a$, where $P_0= X$ and $P_1=Y$.
		A Pauli correction is applied conditioned on the outcome of this ancilla measurement. 
		(e) For physical-level distillation, the multi-qubit Pauli measurement must be broken down further into smaller pieces.
		Up until the dashed line involves the construction of a cat state on neighboring ancilla qubits. Note that if the support of the multi-qubit Z measurement is not over all six qubits, we replace the upcoming ZZ measurement for any qubits not in the support, and replace it by a single Z measurement.
		The Pauli correction at the end depends on the outcomes of the measurements in the circuit.
		\textbf{Logical distillation.}---
		These operations allow the distillation circuit to be re-expressed as surface code operations using 20 tiles to form a logical distillation unit.
		The five blue tiles store the qubits shown in the original circuit, and as such, four of them contain T states at the start of the distillation unit.
		The three green tiles also host T states, and these will be refreshed as the circuit proceeds.
		Each of the diagonal non-Clifford rotations is implemented using the gadget from (d), where the T state is held in one of the green tiles, and the five grey tiles represent ancillas that are used to facilitate the multi-qubit Z measurement, which takes a single logical time step. 
		Following the multi-qubit Z measurement, that green ancilla is measured in either the X (taking zero logical time steps) or the Y basis (taking one logical time step with the aid of three adjacent ancilla tiles). Following this, another T state is transferred to the green tile, which we assume can be done in a single logical time step. 
		As we have three green tiles, we can overlap this procedure for consecutive rotations such that the 11 non-Clifford rotations are each implemented in a total of 13 logical time steps.
		This represents the full time of the distillation unit since the initial X basis preparation and final single-qubit measurements take zero logical time steps.
		\textbf{Physical distillation.}---
		The physical distillation unit can be run on 12 physical qubits laid out as shown. 
		Each of the 11 non-Clifford rotations requires five time steps, but the ancilla preparation in the X basis for all but the first round since the X basis at the end of the previous round deems it unnecessary.
		As the operations at the very start and end of the circuit are not supported on the ancillas, the can be done in parallel to the first and last steps of the non-Clifford rotations, such that this physical distillation unit requires $5 + 10\cdot 4 + 1 = 46$ physical time steps.
		}\label{fig:distillation-circuits-1}
\end{figure}

\begin{figure}[h]
	\centering
	\includegraphics[width=1.0\textwidth]{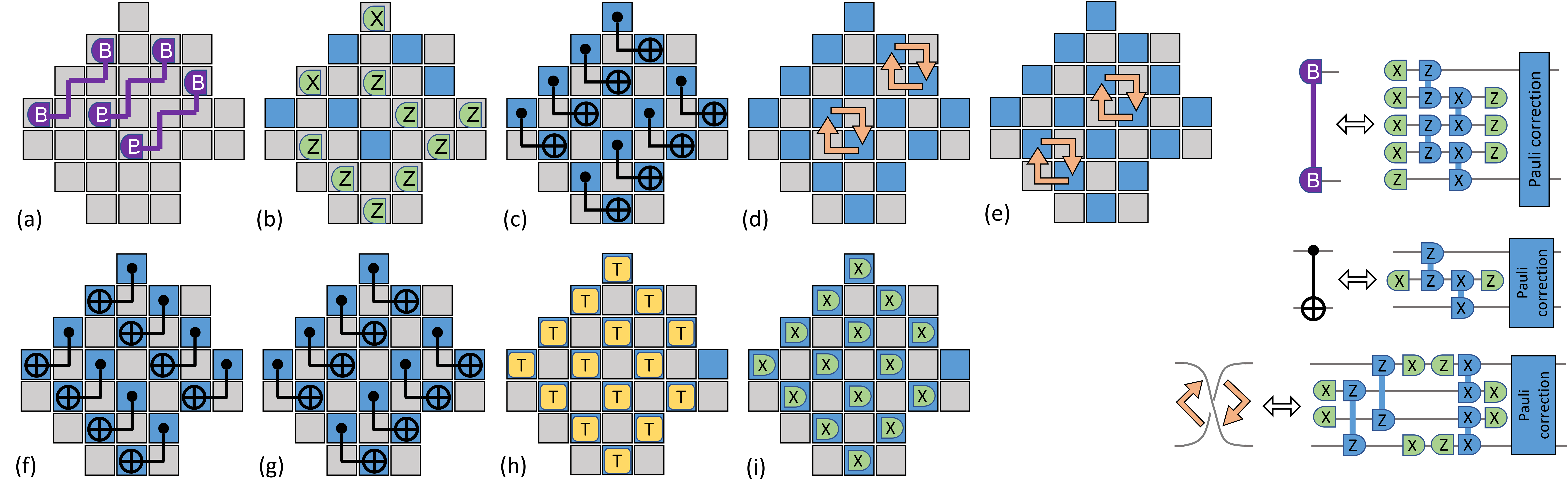}
	\caption{%
		Reed-Muller state preparation 15-to-1 distillation units.
		Each of the 31 squares represents a physical Majorana qubit or a logical qubit in a one-tile patch.
		In steps (a-g), a Reed-Muller state is prepared, using the gadgets shown on the right, which are available for both physical Majorana qubits and logical operations.
		We consider the grey squares ancillas, and the blue squares are those qubits which have been set and acted upon and host the Reed-Muller state.
		A naive counting of these first seven steps requires $(4+1+4+6+6+4+4)=29$ physical time steps, however overlapping consecutive steps with disjoint support reduces this to 21 physical time steps.
		At the logical level, these seven steps require $(2+0+2+2+2+2+2)=12$
		logical time steps.
		(h) Next a T gate is applied to all but one of the qubits in the Reed-Muller state.
		For physical qubits, this can be done directly in a single physical time step.
		For logical qubits, this is instead achieved by transferring a T state to one of the grey qubits, and applying a $ZZ$ measurement as in \fig{distillation-circuits-1}(b). As we do not account for the transfer time, this requires a single logical time step.
		(i) Finally, each of the qubits to which T gates had been applied are measured in the X basis, which takes a single physical time step, or zero logical time steps.
		The overall time for the physical distillation unit is therefore 21+1+1= 23 physical time steps, and for the logical distillation unit is 12+1+0=13 logical time steps.
		}\label{fig:distillation-circuits-2}
\end{figure}

The outputs of each distillation round are fed into the next, which leads to the following sequence of calculations which can be used to identify the overall properties of the factory.
First we set the input error rate $Q_0$ for the first round to be the physical error rate of the T, i.e., $Q_0=p_T$.
Then let $Q_{r-1}$ be the error rate of the encoded T state that is input into round $r$, and let $P_r$ be the Clifford error rate of the unit at that round (where $P_r$ is calculated from the physical qubit type and the distance and type of error correcting code used by the unit).
The acceptance probability $P^\text{acc}_r$ and output T state error rate $Q_{r}$ for each unit in round $r$ are then calculated by setting $P_T$ to $Q_{r-1}$ and setting $P$ to $P_r$.
In \tab{distillation-modules} we show the relevant formulas for the four explicit distillation units specified in \fig{distillation-circuits-1} and \fig{distillation-circuits-2}, which are all different explicit implementations of the standard 15-to-1 protocol~\cite{bravyi2005}.

The overall output T states of a distillation factory $\mathcal{D}$ are those output from its last round, and therefore have error rate 
\begin{eqnarray}
P_T(\mathcal{D}) = Q_R.
\end{eqnarray}
We estimate the run time and qubit requirements as:
\begin{eqnarray}
\tau(\mathcal{D}) & = & \sum_{r=1}^R \tau(M_r),\\
n(\mathcal{D}) & = & \text{max}_{r\in \{ 1,2,\dots R\}}(c_r n(M_r)),
\end{eqnarray}
where $n(M_r)$ is the number of qubits needed for a single copy of distillation unit $M_r$ and $c_r$ is the number of copies of the unit in round $r$. 
Each unit in a factory can fail, therefore the number of T states output by a factory is a random variable.
We choose to address this by defining the number of output T states from the factory $M(\mathcal{D})$ as follows:
\begin{eqnarray}
M(\mathcal{D}) & = & \# ~\text{T states output in at least 99\% of runs of}~ \mathcal{D}.
\end{eqnarray}
In principle, many factories have $M(\mathcal{D}) = 0$ according to this definition, which would occur for example if the final round consists of a single unit with an acceptance probability below 99\%.
Therefore many of the distillation factories that we consider over-provision units so that after accounting for some units failing the required number of T states is passed on to the next round. We assume that such decisions are specified to our framework by setting the appropriate distillation factory parameters such as $R$, $c_r$ and choice of unit types for each round.

\begin{table}[h]
	\begin{tabular}{|c||c|c|c|c||c|c|c|c||c|c|}
		\hline
	    \multirow{2}{*}{$P_T$} & \multicolumn{4}{|c||}{round 1}		& \multicolumn{4}{|c||}{round 2} &  \multicolumn{2}{|c|}{resources} \\
		\cline{2-11}
		& unit & qubits & time & copies	& 	unit & qubits & time & copies  & qubits & time \\
		\hline
		\hline
	    & 15-to-1 & & & & & & & & & \\
		$5.6 \cdot 10^{-11}$ & space-eff. & 3240 & 46.8 $\mu$s & 1 & & & & & 3240 & 46.8 $\mu$s \\
		& $(d=9)$ & & & & & & & & & \\
		\hline
		& 15-to-1 & & & & 15-to-1 & & & & & \\
	    $2.1 \cdot 10^{-15}$ & space-eff. &  $16000 = 16 \cdot 1000$ & 26 $\mu$s & 16 & RM prep. & 10478 & 57.2 $\mu$s & 1 & 16000 & 83.2 $\mu$s \\
		& $(d=5)$ & & & & $(d=13)$ & & & & & \\
		\hline
		& 15-to-1 & & & & 15-to-1 & & & & & \\
		$5.51 \cdot 10^{-13}$ & space-eff. & $5760 = 16 \cdot 360$ & 15.6 $\mu$s & 16 & space-eff. & 4840 & 57.2 $\mu$s & 1 & 5760 & 72.8 $\mu$s \\
		& $(d=3)$ & & & & $(d=11)$ & & & & & \\
		\hline
	\end{tabular}
	\caption{
	T state distillation factories to target the required error rates for our three application examples with the (ns,~$10^{-4}$) qubit parameter example.
    }
	\label{tab:distillation-examples}
\end{table}

\newpage
\section{Compilation from QIR to the planar quantum ISA}
\label{app:cat-state}

The planar quantum ISA, consisting of the Clifford operations defined in \fig{logical-operations} and non-Clifford T states distilled as described in \app{distillation-overhead}, forms a set of basic instructions that is sufficient for universal quantum computation. 
Here we describe how the program expressed in QIR, is mapped to the instructions of the planar quantum ISA, which we refer to as back end compilation. 

Here we describe the back end compilation scheme that we assume in this paper at a high level.
Our compilation scheme is formed by combining sequential Pauli-based computation (SPC), as described in Ref.~\cite{Litinski2019} and elaborated upon in Ref.~\cite{Chamberland2021}, with an approach to synthesize sets of diagonal non-Clifford unitaries in parallel, as was done in Ref.~\cite{Beverland2022}.
We call this compilation scheme Parallel Synthesis Sequential Pauli Computation (PSSPC).

We assume that the input is a QIR program which has first been re-expressed as a sequence of Clifford layers and non-Clifford layers, which we refer to as the input circuit. 
A Clifford layer of the input circuit consists of a set of basic instructions, which are either Clifford unitaries, basis state preparations, or Pauli measurements.
Non-Clifford layers of the input circuit require T states to implement, and include operations such as T gates, Toffoli gates, and single-qubit arbitrary angle rotations (where the angle is not an integer multiple of $\pi/4$) around either the X, Y or Z axis.
The input circuit can be dynamic in the sense that the measurement outcomes in one Clifford layer can affect the instructions applied in subsequent layers of the input circuit.
In the following description, we describe a compilation strategy which can be applied to compile a dynamic input circuit layer by layer\footnote{When estimating resources to implement a dynamic algorithm, we randomly pick outcomes of measurements in the algorithm such that each layer is specified for explicit analysis.} to produce an equivalent output circuit which is expressed in terms of the logical instruction set in \fig{logical-operations}.

\begin{figure}[b]
\centering
\includegraphics[width=0.95\textwidth]{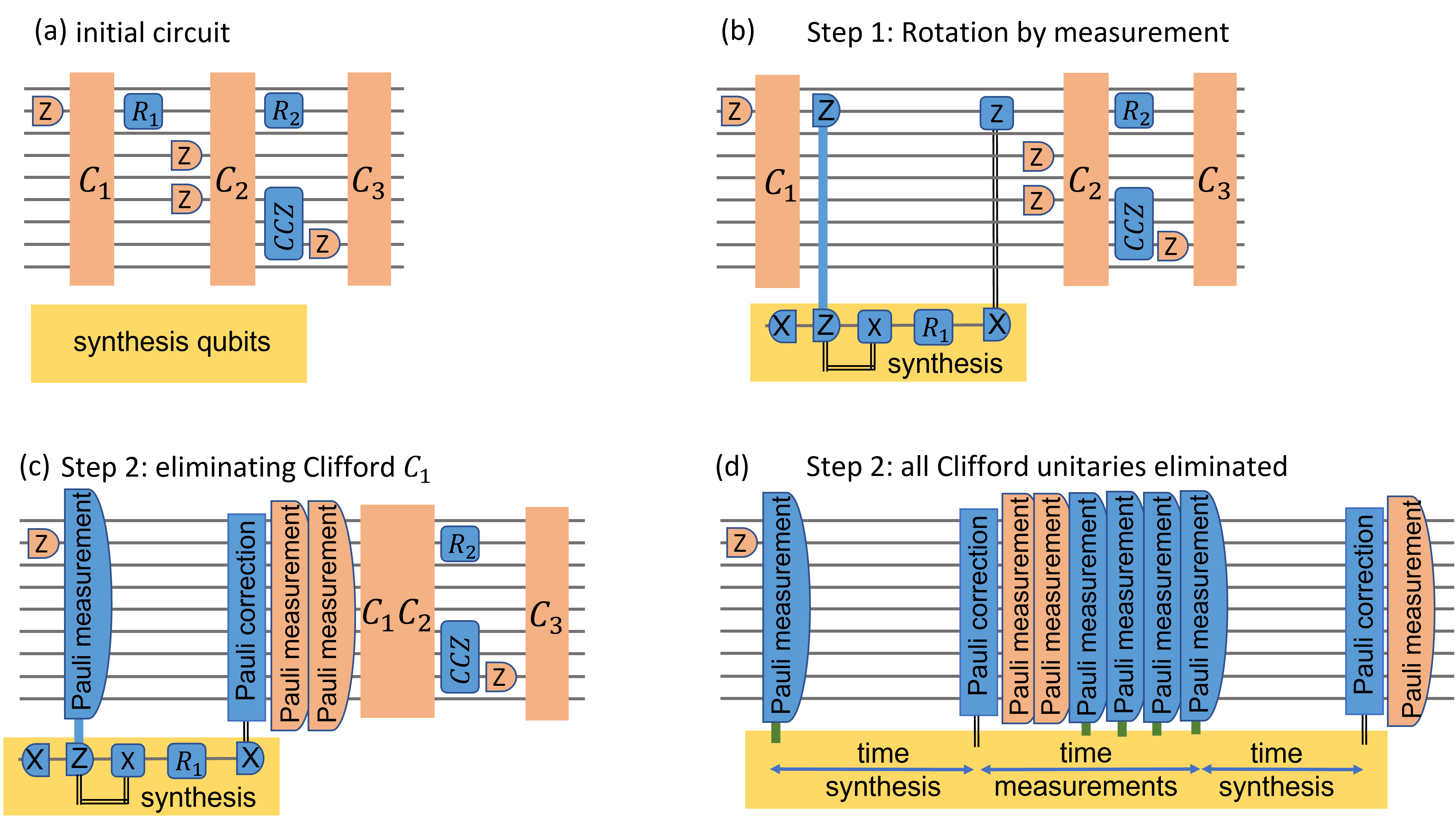}
\caption{
(a) The initial circuit consists of alternating Clifford (orange) and diagonal non-Clifford layers (blue). 
For illustrating the principles behind PSSPC, we show the implementation details for the first non-Clifford (the single rotation $R_1$ between Clifford unitaries $C_1$ and $C_2$), but the concepts extend to parallel distinct rotations. 
(b) $R_1$ is delegated to an ancilla qubit, where the operation is synthesized using T states and the phases are kicked back to the algorithm qubit. 
(c) $C_1$ is eliminated by commuting it through the Pauli operations that follow it, transforming them into multi-qubit Pauli operations. 
(d) After eliminating all Clifford unitaries, the circuit consists of a sequence of Pauli measurements and Pauli corrections, along with synthesis on ancilla qubits.
For the second set of non-Cliffords (between $C_2$ and $C_3$ in the input circuit), we have a series of Pauli measurements and synthesis, followed by a final Pauli correction after synthesis and then the Pauli measurements from the original circuit.
We assume that the Pauli correction for the current non-Clifford layer is completed before moving on to the next layer.
}
\label{fig:sequential-Pauli-computation}
\end{figure}

To orchestrate the execution of this algorithm, PSSPC performs the following circuit transformations (see \fig{sequential-Pauli-computation}).  
\begin{itemize}
    \item[] {\bf Step 0.}---
    We first re-express the input algorithm such that it consists of a sequence of Clifford layers and diagonal non-Clifford layers. 
    This is straight-forward: a rotation around the X or Y axis can be converted by conjugation of a Clifford into a rotation around the Z axis $R_z(\theta) = \ket{0}\bra{0} + e^{i \theta} \ket{1} \bra{1}$.
    The Cliffords that are used to implement this transformation are then absorbed into the Clifford layers before and after the non-Clifford layer.
    Similarly, a Toffoli gate can be re-expressed as a CCZ gate with a Hadamard applied before and after on the target qubit.

    \item[] {\bf Step 1.}---
    Delegating expensive operations to synthesis qubits to increase parallelism: Non-Clifford rotations $R_z(\theta)$ are frequently used in quantum algorithms, but are among the most resource intensive operations when we consider implementation costs. 
    To improve algorithm performance, PSSPC seeks to delegate these operations to ancilla qubits, referred to as synthesis qubits. 
    By delegating these operations, multiple rotations can be synthesized in parallel to increase the T consumption rate and improve algorithm runtime. 
    
    As shown in \fig{sequential-Pauli-computation}(b), PSSPC first re-expresses rotation operations on algorithm qubits using ancillary synthesis qubits. 
    To implement a rotation such as $R_z(\theta)$ on an algorithm qubit, it is first entangled with an ancilla by measuring a Pauli supported on both that qubit and the ancilla.
    Then a sequence of operations are performed on the ancilla to apply the rotation's phase using Clifford gates and also T gates. 
    The phase is kicked back to the algorithm qubit by measuring out the ancilla and applying a Pauli correction to the algorithm qubit.
    All other diagonal non-Clifford unitaries (including the T gate and the CCZ) are applied similarly~(see Figure~20 in~\cite{Beverland2022}).

    \item[] {\bf Step 2.}---
    Eliminating Clifford unitaries:
    After delegating the synthesis of rotations and Toffoli gates to ancillas, operations on the algorithm qubits are a series of Clifford unitaries and Pauli measurements. 
    While these operations can be implemented using the instruction set in \fig{logical-operations}, we can further optimize the circuit by eliminating Clifford unitaries. 
    
    Clifford unitaries map Pauli operations to other Pauli operations. 
    Using this property, PSSPC commutes each Clifford unitary through the subsequent Pauli operations. 
    This transforms the local Pauli operations to multi-qubit Pauli measurements and Pauli corrections. 
    This optimization allows us eliminate the Cliffords at the cost of executing multi-qubit Pauli measurements.
\end{itemize}

Like SPC, a major benefit of PSSPC is that all Clifford unitaries in the input algorithm are eliminated and do not need to be applied.
SPC differs from PSSPC by synthesizing the non-Clifford gates in place rather than in external ancillas, and the Ts  resulting algorithm which has the drawback of serializing all T gates in the circuit. 
This drawback is partially overcome by PSSPC by parallelizing the synthesis of rotations which appear in the same layer of the input circuit, but the phase kickback of the synthesized rotations within a layer is still serial.
Other layout and compilation approaches have been proposed~\cite{JavadiAbhari2017,Lao2018,Steiger2018,Murali2019,Zulehner2019,Childs2019,Beverland2022,Leung2010,Hahn2019,Beaudrap2020a}, but are more involved to implement and analyze in detail.

We estimate the resources required to implement an algorithm using PSSPC in terms of the following parameters of the input algorithm. Let $\Qalg$ be the number of logical qubits used by the algorithm. 
Let $\RU$ be the number of single-qubit rotations, $\RT$ be the number of T gates, and $\RTof$ be the number of Toffoli gates, and $\RMeas$ be the number of Pauli measurements of the input algorithm.
Let $\depthU$ be the number of non-Clifford layers in which there is at least one arbitrary angle rotation (note this can be smaller than the total number of non-Clifford layers since it excludes those layers consisting entirely of T gates and Toffoli gates). 
Let the target error budget for synthesis in the overall program be $\epsSynth$ (measured in the diamond norm). 

\begin{figure}[b]
\centering
\includegraphics[width=.85\textwidth]{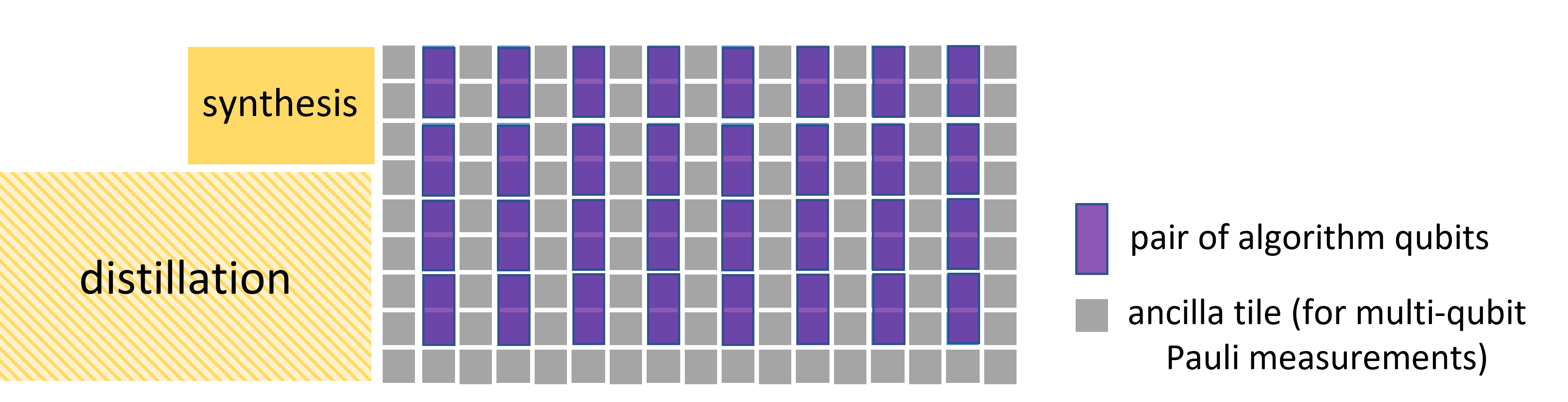}
\caption{
Overall layout of qubits in the quantum computer, using the PSSPC layout. 
The main block of logical qubits is purple and grey. 
Each purple rectangle denotes a pair of algorithm qubits encoded in a two-tile patch of surface or Hastings-Haah code, while the grey boxes are tiles of ancillas to facilitate multi-qubit Pauli measurements. 
A block of qubits is dedicated for distilling T states which are used to apply T gates and Toffolis, or are passed on to the synthesis block to synthesize rotations. 
We neglect the qubits needed for synthesis.
In the overall program, we use $\Q$ to denote the number of tiles in the main block, $\R$ to denote the number of T states that are produced, and $\TimeLogical$ to denote the number of logical time steps needed to run the program.
}
\label{fig:compilation-cat-state-layout}
\end{figure}

We assume the fast block layout from Ref.~\cite{Litinski2019} (also shown in \fig{compilation-cat-state-layout}), such that each pair of algorithm qubits is stored in a two-tile surface code patch, surrounded by ancilla tiles which can be used to measure an arbitrary Pauli operator on the algorithm qubits in one logical time step.
This results in the number of logical qubits\footnote{More precisely, $\Q$ is the number of tiles, some of which are allocated to ancillas, and some of which are allocated to store algorithm qubits in two-tile patches.} $\Q$ given by
\begin{equation}
\label{eq:logical-qubits}
\Q = 2\Qalg + \left\lceil \sqrt{8\Qalg} \right\rceil + 1.
\end{equation}
We account for the resources required for distillation separately in \app{resource-estimates}, and in this section we assume that T states are produced on command and simply count how many T states are needed to implement the algorithm.

Synthesizing a single-qubit rotation to diamond-norm accuracy $\epsilon'$ re-expresses the unitary as Cliffords and at most $\TperU(\epsilon')$ non-Clifford T gates, where
\begin{eqnarray}
\label{eq:T-count-synthesis}
\TperU(\epsilon') = \left\lceil A\log_2(1/\epsilon')+B \right\rceil,
\end{eqnarray}
for constants $A$ and $B$.
We assume the Clifford+T synthesis in Table 1 of Ref.~\cite{Kliuchnikov2022}, which results in $A=0.53$ and $B=5.3$.\footnote{For simplicity, we use this formula for all single-qubit arbitrary angle rotations, and do not distinguish between best, worst and average cases.
We also do not incorporate further improvements from~\cite{Kliuchnikov2022} that use Clifford+$\sqrt{\mathrm{T}}$ synthesis and known improvements for special cases, for example when many rotations are applied in parallel with the same angle~\cite{Gidney2018halvingcostof}.}
To meet the accuracy requirement for the overall algorithm, which involves $\RU$ single-qubit rotations, we set $\epsilon'=\epsSynth/\RU$.

As described in \fig{sequential-Pauli-computation}(d), the computation proceeds in layers.
Each diagonal non-Clifford unitary is applied as part of a layer of non-Clifford rotations from the input circuit.
The first step to implement such a unitary is to couple the algorithm qubits to the synthesis ancilla qubits.
This requires one multi-qubit Pauli measurement for each algorithm qubit in the support of the unitary.
Therefore one logical time step is used for a T gate or a single-qubit rotation $R_z(\theta)$, while each Toffoli requires three logical time steps.
We assume that T gates and CCZ gates require no additional synthesis time beyond the time required for these measurements, justified by techniques in Ref.~\cite{Gidney2019,Haner2022}.
To estimate the time required to synthesize an arbitrary angle rotation $R_z(\theta)$, we note that the synthesis consists of $\TperU(\epsSynth/\RU)$ non-Clifford T gates interleaved with Clifford unitaries. 
We assume this can be implemented by a sequence of $\TperU(\epsSynth/\RU)$ Pauli measurements supported on the synthesis ancilla and other ancillas to which the T states are delivered from distillation factories, taking $\TperU(\epsSynth/\RU)$ logical time steps.
The final step to implement a rotation, after the synthesis has finished, is to apply a measurement of the ancilla and then a Pauli correction to the algorithm qubits, both of which can be done instantaneously.

The time required to implement a non-Clifford layer depends on whether the layer consists entirely of Toffoli and T gates, or whether it also contains some arbitrary angle rotations.
In the former case, the number of logical time steps required is simply the number of measurements needed to entangle the algorithm and synthesis ancillas.
In the latter case, in addition to the time needed for these entangling measurements, one also needs to add the time for synthesis.
It is possible to further compress this by overlapping synthesis for one non-Clifford layer with the measurements for the following Clifford and non-Clifford layers, but for simplicity we assume that consecutive layers are executed sequentially without overlap.
Within each layer however, the synthesis of rotations that were parallel in the input quantum IR circuit is performed in parallel.
Since all Clifford unitaries in the Clifford layer are eliminated, the time required for the Clifford layer is simply the time required to implement the measurements it contains.
The total number $\TimeLogicalFast$  of logical time steps is therefore
\begin{equation}
\label{eq:logical-timesteps}
\TimeLogicalFast =  (\RMeas + \RU +\RT) + \left\lceil A \log_2(\RU/\epsSynth) + B \right\rceil \depthU   + 3\RTof .
\end{equation}
Note that this is our estimate of the number of logical time steps that would be needed if the T state factories produce T states at a sufficient rate so as to not limit the algorithm speed. 
In practice, the time $\TimeLogical \geq \TimeLogicalFast$, and one could increase $\TimeLogical$ to slow the algorithm down if for example it puts a strain on the qubit resources available to produce T states.

Finally, using \eq{T-count-synthesis} and noting that four T states are sufficient to implement Toffoli,
the number of T states $\R$ is given by
\begin{equation}
\label{eq:number-T-states}
\R = \left\lceil A \log_2(\RU/\epsSynth) + B \right\rceil\RU + 4\RTof + \RT.
\end{equation}

We neglect the ancilla qubits used for unitary synthesis and those that would be required to transport the T states from distillation factories to where they are consumed for synthesis.

\section{Putting the pieces together to estimate resources of an algorithm}
\label{app:resource-estimates}

This appendix puts together components described over previous appendices in order to estimate the resources required to implement a high-level language quantum algorithm on a specified fault-tolerant architecture. 
After compilation of the algorithm to the instruction set described in \fig{logical-operations}, the problem reduces to estimating the resources required to implement a circuit with $\Qalg$ algorithm qubits, with $\RU$ single-qubit rotations spread among $\depthU$ layers, $\RT$ T gates, and $\RTof$ Toffoli gates.
Furthermore, let $1-\epsTot$ be the desired probability that the overall algorithm succeeds.\footnote{
More formally, $\epsTot$ is the total variational distance between the probability distribution of the final output bit strings of the compiled circuit and the algorithm circuit. 
Commonly, the circuits correspond to channels and the required total variational distance can be bounded using the diamond norm distance between the compiled channel and algorithm's channel as explained in Appendix~B in~\cite{Kliuchnikov2022}.
}

The value of $\epsTot$, which we refer to as the \textit{error budget}, depends on how the samples from algorithm instances are to be post-processed, and its value is set by taking the desired algorithm execution accuracy to be $1-\epsTot$.
For example, if one is running Shor's algorithm for factoring integers, a large value of $\epsTot$ may be tolerated as one can check that the output are indeed the prime factors of the input.
On the other hand, a much smaller $\epsTot$ may be needed for an algorithm solving a problem with a solution which cannot be efficiently verified.
For the target architecture, we assume that a specific physical qubit model and quantum error correction scheme has been chosen from \app{noise} and \app{logical-operations-details}. The target architecture implements the logical instruction set described in \fig{logical-operations}.

To achieve the required algorithm success probability, we require that 
\begin{equation}
\label{eq:error-budget}
  \epsLog +\epsDis+\epsSynth \leq \epsTot,
\end{equation}
where $\epsLog$, $\epsDis$ and $\epsSynth$ are the probabilities of at least one logical failure, at least one faulty T distillation, and at least one failed rotation synthesis respectively.
In our estimates, we retain only the first order terms in these small probabilities.
We ensure \eq{error-budget} is satisfied by requiring that each of $\epsLog$, $\epsDis$ and $\epsSynth$ are at most $\epsTot /3$.

We assume the PSSPC compilation strategy from \app{cat-state}, which sets the number of logical qubits $\Q$ and the number of T states consumed $\R$.
This also specifies a minimum possible logical runtime of the algorithm $\TimeLogicalFast$.
In what follows we establish the code distance $d$ required for error correction, the choice of distillation factory $\mathcal{D}$, the number of distillation factories $\nFactories$, and the number of logical time steps $\TimeLogical$.
We then use these to identify the number of physical qubits $\qphy$ and runtime $\tphy$. 

Before computing these quantities, let's consider the need to balance the speed of the algorithm vs. the rate of T state production. 
From \app{cat-state}, we know that the algorithm requires at least $\TimeLogicalFast$ time steps. That is, we are free to run the algorithm at any desired speed by choosing $\TimeLogical \ge \TimeLogicalFast$. 
Similarly, we are allowed to choose the number of factories $\nFactories$ and control how many T states are available for the algorithm per logical time step. 
However, we require that for any chosen value of $\TimeLogical$, $\nFactories$ must be chosen large enough to produce all the required T states for the algorithm within $\TimeLogical$ time steps. Therefore, with large $\TimeLogical$, as we slow down the algorithm, we can afford to reduce the number of T factories and save qubits required for their implementation\footnote{Note that slowing down the algorithm may not always result in a reduction in overall qubit counts. As the algorithm is slowed down, we can use lesser T factories and incur less qubit overhead for them. But for the algorithm's logical qubits, the code distance depends on the number of logical time steps (see \eq{distance-required}). We may require a higher code distance to protect the algorithm qubits during a slower execution}. Similarly, as we speed up the algorithm to finish in the least number of logical time steps possible, we require a large number of T factories to quickly supply the required T states, and incur higher qubit overheads. To navigate this space-time tradeoff, we fix a scenario such as space-optimal or time-optimal or an intermediate case and choose suitable values for $\TimeLogical$ and $\nFactories$. 
In what follows, we present the analysis for the general case where some $\TimeLogical \geq \TimeLogicalFast$ has been selected. 

{\noindent \textbf{Estimating resources for logical operations:}} We determine the number of physical qubits and time required for the logical operations by selecting an appropriate code distance used for the logical qubits. 
From \eq{logical-error-rate-formula}, we see that $d = 2 \log (a/\Plog)/\log(p^*/p)-1$ for constants $a$ and $p^*$ that depend on the type of physical qubit chosen in the architecture.
Seeking the smallest error correction overhead that achieves an acceptably low probability that logical failure occurs in the implementation of the algorithm, i.e., that $\Q \TimeLogical \Plog = \epsLog \leq \epsTot/3$, we select the following distance,
\begin{eqnarray}
\label{eq:distance-required}
d = \left\lceil
\frac{2\log(a\epsTot/(3QC))}{\log(p^*/p)}
-1\right\rceil_\text{odd}.
\end{eqnarray}
We then obtain the physical run time by multiplying the number of logical time steps $\TimeLogical$ by the time per logical time step $\tau(d)$ as defined in \app{logical-operations-details},
\begin{eqnarray}
\tphy=\tau(d) \TimeLogical.
\label{eq:logical-time}
\end{eqnarray}
Similarly, the number of physical qubits required for the logical operations is $\Q n(d)$.
Note that the code distance is a function of $\TimeLogical$. If $\TimeLogical$ is chosen to be much higher than $\TimeLogical_{min}$ to reduce the number of factories, we may require a higher code distance to protect the algorithm qubits for the longer execution duration.

{\noindent \textbf{Estimating resources for T distillation:}}
Next, we determine $\mathcal{D}$, the number of factories and the physical resources required for distillation.
Towards this, we first require the number of T states required by the algorithm at the given error budget. 
That is, we need to satisfy \eq{number-T-states} with $\epsSynth \leq \epsTot/3$, which allows us to calculate $\R$, the minimum number of T states required.
Next, we determine the quality of these T states.
In \tab{hardware-examples}, we show the number of physical qubits $n(\mathcal{D})$ and the time $\tau(\mathcal{D})$ to distill a single T state with error rate $\PlogDis$ with various choices of distillation factory $\mathcal{D}$.
As we aim for the $\R$ injected T states to fail with low probability, we require that $\R \PlogDis(\mathcal{D}) = \epsDis \leq \frac{\epsTot}{3}$.
Given a list of all the considered factories with the specified hardware type $\{ \mathcal{D}_1, \mathcal{D}_2,  \mathcal{D}_3, \dots,  \mathcal{D}_m \}$, we therefore select $\mathcal{D}$ as the factory with the smallest space-time footprint which satisfies this inequality, i.e.,
\begin{eqnarray}
\label{eq:select-factory}
\mathcal{D} = \argmin_{\mathcal{D}_i \in \{ \mathcal{D}_1, \mathcal{D}_2,  \mathcal{D}_3, \dots,  \mathcal{D}_m \}}{\{ n(\mathcal{D}_i)\tau(\mathcal{D}_i) | \PlogDis(\mathcal{D}_i) \leq \epsTot/3\R\}}.
\end{eqnarray}

Finally, the smallest number $\nFactories$ of distillation factories capable of producing the demanded $\R$ T states during the algorithm's run time is selected
\begin{eqnarray}
\nFactories = \left\lceil \frac{\R \cdot \tau(\mathcal{D})}{M(\mathcal{D}) \cdot \tphy} \right\rceil.
\end{eqnarray} 

At this point a check should be made to ensure that the factories execute during the runtime of the algorithm, i.e., that $\tau(d) \TimeLogical \geq \tau(\mathcal{D})$.
This catches the case when the runtime of an algorithm is less than the runtime of a single T factory.
If this is not satisfied, $\TimeLogical$ should be increased and the analysis repeated.

{\noindent \textbf{Total resources:}}
The total number of qubits $\qphy$ for the quantum computation is then found from adding those qubits for distillation to those required to build the logical qubits,
\begin{eqnarray}
\label{eq:total-physical-qubits}
\qphy = \nFactories n(\mathcal{D}) + \Q n(d).
\end{eqnarray}

The total runtime of the algorithm is given by \eq{logical-time}.

\section{Applications}
\label{app:applications}

Here we provide further details of the three example applications, and the calculations that result in the resource estimates in \tab{physical-and-logical-overhead}.

Before getting into specifics about each of three application examples, we make a remark about quantum algorithm sampling.
Throughout this work, we are focused on the resources required to implement a full quantum algorithm once, however it is often the case that quantum algorithms must be repeated (even when run perfectly). 
This is because the output of some quantum algorithms is not a deterministic bit string, but instead is a drawn from a probability distribution over bit strings, and the information from the algorithm may require learning details of the distribution, requiring a set of samples.
We do not account for the resource costs that algorithms may need to be sampled many times in this work. 
This could in principle be achieved by either re-running the algorithm consecutively on the available quantum computer, or by having many independent quantum computers running in parallel, or some combination.

{\bf Quantum dynamics}.---
Here we outline the calculation that produced the resource requirements of the quantum dynamics example in \tab{physical-and-logical-overhead}, following the approach in Ref.~\cite{Pearson2020}.
We are interested in the time evolution $e^{-iHt}$ by the 2D $\sqrt{N}\times \sqrt{N}$ transverse field Ising Hamiltonian
\begin{align}
H=\underbrace{-J\sum_{\langle j, k\rangle} Z_jZ_k}_{A}+\underbrace{g\sum_{j}X_j}_{B}    
\end{align}
with nearest-neighbour interactions is accomplished using fourth-order product formulas. 
Trotter \& Suzuki~\cite{Hatano2005TrotterSuzuki} define a recursive construction using
\begin{align}
\label{eq:TrotterFourthOrder}
    U_2(\Delta)&= e^{-iA\Delta/2}e^{-iB\Delta}e^{-iA\Delta/2}=e^{-iH\Delta}+\mathcal{O}(\Delta^3),\\
    U_4(\Delta)&=U_2(\gamma \Delta)U_2(\gamma\Delta)U_2((1-4\gamma)\Delta)U_2(\gamma \Delta)U_2( \gamma\Delta)=e^{-iH\Delta}+\mathcal{O}(\Delta^5),\nonumber\\
    \gamma &=(4-4^{1/3})^{-1}.\nonumber
\end{align}
When applying $T>1$ time steps steps, the first and last terms may be merged, leading to $10T+1$ exponentials in~\eq{TrotterFourthOrder}. 
Note that $A$ and $B$ may be interchanged depending on which is more expensive. 
These product formulas reduce the resource costs of quantum dynamics to evaluating the resource cost of applying $e^{-iA\Delta }$ and $e^{-iB\Delta }$. 
All terms in $A$ commute and similarly for $B$, and $e^{-iB\Delta }$ is the product of $N$ single-qubit rotations $\prod_{j}e^{-iX_jg\Delta}$. 
The simulation can be carried out by assigning a qubit to each of the $N$ sites in the lattice, such that $\Qalg=N$.
By conjugating with a CNOT gate, each nearest-neighbour $Z_jZ_k$ term is transformed in to a single-qubit Pauli $Z_j$. 
Thus $e^{-iA\Delta }$ can be implemented by eight depth-1 layers of $N/2$ CNOT gates and four depth-1 layers of $N/2$ single qubit rotations $\prod_{j}e^{iZ_jJ\Delta}$.
The number of rotations is therefore $\RU = (5T+1) N+10T N= (15T+1) N$. 
Finally, the depth of the rotation gates over the $T$ time steps is $\depthU = 5T + 1 +4(5T)=25T+1$.

In the example we consider, we take $\epsilon = 0.001$, $N=100$ and $T=20$ such that when the algorithm, which is initially written as a Q\# program, is explicitly compiled down to a QIR-level program with $\Qalg=100$, $\RU=30100$, $\RT=0$, $\RTof=0$ and $\RMeas= 1400000$ and $\depthU=501$.
From \eq{logical-qubits}, \eq{logical-timesteps} and \eq{number-T-states}, we calculate that with the PSSPC compilation and layout, the ISA-level executable that this QIR program would compile down to has $\Q = 230$ logical qubits, which are needed for at least $\TimeLogicalFast = 1.5 \cdot 10^{5}$ logical time steps, consuming $\R = 2.4 \cdot 10^{6}$ T states.

At the physical level, here we illustrate the analysis with the (ns,~$10^{-4}$) qubit example (with the other qubit parameter examples being analyzed analogously).
In this example we will ultimately increase $\TimeLogical$ so that it is significantly larger than $\TimeLogicalFast$ to reduce the qubit count as a trade-off paid for by an increase in run time. 
To see why we do so, let us first consider the case where $\TimeLogical = \TimeLogicalFast$.
We use \eq{distance-required} to find the distance $d = 9$, such that $n(d) = 2 d^2 = 162$ and $\tau(d) = (4 \tGate + 2 \tMeas) d =400$ ns.
This sets the algorithm run time via \eq{logical-time} to be $\tphy =$ 0.55 s.
From \eq{select-factory}, a distillation factory is selected.
There are $\nFactories = 199$ factories. 
Each factory has a single round of logical distillation (15-to-1 space efficient), with one unit of code distance 9.
Each factory produces $M(\mathcal{D}) = 1$ encoded T state with error rate $5.6 \cdot 10^{-11}$, and contains $n(\mathcal{D}) = 3240$ physical qubits, and requires $\tau(\mathcal{D}) = 46.8~\mu$s to run.
The total number of physical qubits using is then \eq{total-physical-qubits} is $\qphy = $ 0.68M, with distillation accounting for $94.5 \%$ of these qubits.

Because of the very high fraction of qubits allocated to distillation when $\TimeLogical=\TimeLogicalFast$, we consider a slowed down version of the algorithm where we set $\TimeLogical=10\TimeLogicalFast$, therefore easing the requirement on the number of T factories required. 
With the (ns,~$10^{-4}$) qubit example, we use \eq{distance-required} to find the distance increases to $d = 11$ since the logical qubits need to be protected for longer. The number of physical qubits per logical qubit is therefore $n(d) = 2 d^2 = 242$ and $\tau(d) = (4 \tGate + 2 \tMeas) d = 4.4~\mu$s.
This increases the algorithm run time via \eq{logical-time} to be $\tphy =$ 6.78 s.
Precisely the same factory is selected, namely a single round of logical distillation (15-to-1 space efficient), with one unit of code distance 9.
Each factory produces $M(\mathcal{D}) = 1$ encoded T state with error rate $5.6 \cdot 10^{-11}$, and contains $n(\mathcal{D}) = 3240$ physical qubits, and requires $\tau(\mathcal{D}) = 46.8~\mu$s to run.
However, now only $\nFactories = 17$ factories are required.
The total number of physical qubits using is then \eq{total-physical-qubits} drops to $\qphy = $ 0.11M, with distillation accounting for $49.7 \%$ of these qubits.

{\bf Quantum chemistry}.---
We assume the so-called `double-factorized qubitization' algorithm described in Ref.~\cite{vonBurg2021,Low2022Halving} is used. 
The \emph{qubitization} approach is based on quantum phase estimation, but instead of constructing the standard $U = \exp{(-i H/\alpha)}$ from the Hamiltonian matrix $H$, one instead takes $U \approx \exp{(-i \sin^{-1} (H/\alpha))}$, which can typically be implemented with fewer resources by qubitization~\cite{Low2019Qubitization}.
Using \emph{double-factorization}, the naive representation of $H$ is compressed into fewer terms and also with a smaller $\alpha$.

We take $\epsTot = 0.01$.
The algorithm, initially expressed as a Rust program, is explicitly compiled down to a QIR-level program with $\Qalg = 1318$ logical qubits with $\RT = 5.53 \cdot 10^{7}$, $\depthU = 2.05 \cdot 10^{8}$, $\RU = 2.06 \cdot 10^{8}$, $\RTof = 1.35 \cdot 10^{11}$ and $\RMeas = 1.37 \cdot 10^{9}$.
From \eq{logical-qubits}, \eq{logical-timesteps} and \eq{number-T-states}, we calculate that with the PSSPC compilation and layout, the ISA-level executable that this QIR program would compile down to has $\Q = 2740$ logical qubits, which are needed for at least $\TimeLogicalFast = 4.10 \cdot 10^{11}$ logical time steps, consuming $\R = 5.44 \cdot 10^{11}$ T states.
We set $\TimeLogical = \TimeLogicalFast$.

At the physical level, here we illustrate the analysis with the (ns,~$10^{-4}$) qubit example (with the other qubit parameter examples being analyzed analogously).
We use \eq{distance-required} to find the distance $d = 17$, such that $n(d) = 578$ and $\tau(d) = 6.8$ $\mu$s.
This sets the algorithm run time via \eq{logical-time} to be $\tphy =$ 1 month and one day.
From \eq{select-factory}, the following distillation factory is selected:
There are $\nFactories = 17$ factories. 
Each factory has two rounds of logical distillation, with 16 units of code distance 5 (15-to-1 space-eff.) in the first round and 1 unit of code distance 13 (15-to-1 RM prep.) in the second round.
Each factory requires $n(\mathcal{D}) = 16000$ physical qubits, $\tau(\mathcal{D}) =$ 83.2 $\mu$s run time, and produces $M(\mathcal{D}) = 1$ encoded T states with error rate $2.13 \cdot 10^{-15}$.
The total number of physical qubits using \eq{total-physical-qubits} is $\qphy = 1.86$M.

{\bf Factoring}.---
We use a custom optimized implementation of Shor's algorithm~\cite{Shor1994}.
We follow the logical algorithm implementation in Ref.~\cite{Gidney2021}, applying all of their optimizations (including coset representation, windowing over exponents and multiplicands, iterative phase estimation), except that we do not implement carry runways.
As an input number to factorize, we use the RSA-2048 number from the RSA factoring challenge~\cite{Kaliski1991}, and set the trial generator to 7.
Further, we set both the windowing parameters $c_{\text{mul}}$ and $c_{\text{exp}}$ to 5.

First note that by the verifiable nature of the factoring problem, namely that once a solution has been found it can be easily verified, we can tolerate a large probability of the algorithm failing due to a logical fault since if the output is not valid the algorithm can be re-run. 
For this application example, we take $\epsTot = 1/3$.

In the example we consider of factoring a $2048$-bit integer, the algorithm is initially expressed as a Rust program, and is explicitly compiled down to a QIR-level program with $\Qalg = 12581$ algorithm qubits.
Moreover, we find that $\RT = 12$, $\depthU = 12$, $\RU = 12$, $\RTof = 3.73\cdot 10^{10}$ and $\RMeas = 1.08 \cdot 10^{9}$.
From \eq{logical-qubits}, \eq{logical-timesteps} and \eq{number-T-states}, we calculate that with the PSSPC compilation and layout, the ISA-level executable that this QIR program would compile down to has $\Q = 25481$ logical qubits, which are needed for at least $\TimeLogicalFast =  1.23 \cdot 10^{10}$ logical time steps, consuming $\R = 1.49 \cdot 10^{10}$ T states.

At the physical level, here we illustrate the analysis with the (ns,~$10^{-4}$) qubit example (with the other qubit parameter examples being analyzed analogously).
We use \eq{distance-required} to find the distance $d = 13$, such that $n(d) = 2 d^2 = 338$ and $\tau(d) = (4 \tGate + 2 \tMeas) d =5.2$ $\mu$s.
This sets the algorithm run time via \eq{logical-time} to be $\tphy =$ 17 hours 43 mins.
From \eq{select-factory}, a distillation factory is selected.
There are $\nFactories = 18$ factories. 
Each factory has two rounds of logical distillation (both 15-to-1 space efficient), with 16 units of code distance 3 in the first round and 1 unit of code distance 11 in the second round, which produces $M(\mathcal{D}) = 1$ encoded T state with error rate $5.51 \cdot 10^{-13}$, and which contains $n(\mathcal{D}) = 5760$ physical qubits, and requires $\tau(\mathcal{D}) = $ 72.8 $\mu$s to run.
The total number of physical qubits using \eq{total-physical-qubits} is then $\qphy =$ 8.72M.

In the introduction, we state that factoring a 2048-bit integer using Shor's algorithm could be done in minutes with an array of twenty five thousand \textit{perfect, noiseless} qubits.
This estimate is obtained by assuming that the perfect qubits replace the ISA-level logical qubits in \tab{application-examples}.
With perfect operations, there would be no need for distillation, and so the only cost would be the qubits $25481$ algorithm and ancilla qubits in the array. 
Taking the $1.2\cdot 10^{10}$ logical time steps to be physical time steps, each lasting 100 ns, results in a run time of 20 mins.

\section{Primary assumptions}
\label{app:primary-assumptions}

Here we summarize our primary assumptions.
These are the subset of assumptions which are not justified to our satisfaction in the current literature, and which we believe will significantly increase our resource estimates if relaxed. 
\begin{enumerate}
    \item[] {\bf Uniform independent physical noise}.---
    We assume that the noise on physical qubits and physical qubit operations is the standard circuit noise model.
    In particular we assume error events at different space-time locations are independent and that error rates are uniform across the system in time and space. 
    \item[] {\bf Efficient classical computation}.---
    We assume that classical overhead (compilation, control, feedback, readout, decoding, etc.) does not dominate the overall cost of implementing the full quantum algorithm.
    \item[] {\bf Syndrome measurement circuits for planar quantum ISA}.---
    We assume that syndrome measurement circuits with similar depth and error correction performance to those for standard surface and Hastings-Haah code patches can be constructed to implement all operations of the planar quantum ISA (as specified in \fig{logical-operations}).
    \item[] {\bf Uniform independent logical noise}.---
    We assume that the error rate of a logical operation is approximately equal to its space-time volume (the number of tiles multiplied by the number of logical time steps) multiplied by the error rate of a logical qubit in a standard one-tile patch in one logical time step.
    \item[] {\bf Negligible Clifford costs for synthesis}.---
    We assume that the space overhead for synthesis and space and time overhead for transport of T states within T state factories and to synthesis qubits are all negligible.
    \item[] {\bf Smooth T state consumption rate}.---
    We assume that the rate of T state consumption throughout the compiled algorithm is almost constant, or can be made almost constant without significantly increasing the number of logical time steps for the algorithm.
\end{enumerate}

\bibliographystyle{plain}
\bibliography{references}
\end{document}